\documentclass[journal,comsoc]{IEEEtran}

\usepackage[T1]{fontenc}% optional T1 font encoding
\usepackage{cite}

\ifCLASSINFOpdf
  \usepackage[pdftex]{graphicx}

\else

  \usepackage[dvips]{graphicx}

\fi
\usepackage{graphicx}
\usepackage{makecell}
\usepackage{amsmath,amsthm,amssymb}
\usepackage{epstopdf}
\usepackage{url}
\usepackage{cite}
\usepackage{color}
\ifCLASSOPTIONcompsoc
  \usepackage[caption=false,font=normalsize,labelfont=sf,textfont=sf]{subfig}
\else
  \usepackage[caption=false,font=footnotesize]{subfig}
\fi

\usepackage{stfloats}

\fnbelowfloat

\begin{document}

\title{Robust Directional Modulation Design for Secrecy Rate Maximization in Multi-User Networks}

\author{Linqing~Gui, ~Mengxia~Yang, ~Xiaobo~Zhou, ~Feng~Shu, ~Jun~Li,

~Jiangzhou~Wang, ~\IEEEmembership{Fellow,~IEEE}, and ~Jinhui~Lu
% <-this % stops a space   ~\IEEEmembership{Member,~IEEE
\thanks{This work is supported in part by the National Natural Science Foundation of China (Nos. 61602245, 61771244, 61501238, 61702258, and 61472190), in part by the Natural Science Foundation of Jiangsu Province (No. BK20150791).}
\thanks{L. Gui, M. Yang, X. Zhou, F. Shu, J. Li and J. Lu are with the Department of Electronic and Optical Engineering, Nanjing University of Science and Technology, Nanjing, 210094 China. E-mail: guilinqing@163.com, 597990359@qq.com, zxb@njust.edu.cn, shufeng0101@163.com, jun.li@njust.edu.cn, jinhui.lu@njust.edu.cn.}
% <-this % stops a space
\thanks{J. Wang is with the School of Engineering and Digital Arts, University of Kent, Canterbury CT2 7NT, U.K. E-mail: j.z.wang@kent.ac.uk.}
}

\maketitle
\begin{abstract}
In this paper, based on directional modulation (DM), robust beamforming matrix design for sum secrecy rate maximization is investigated in multi-user systems.
The base station (BS) is assumed to have the imperfect knowledge of the direction angle toward
each eavesdropper, with the estimation error following the Von Mises distribution.
To this end, a Von Mises distribution-Sum Secrecy Rate Maximization (VMD-SSRM) method is proposed to maximize the sum secrecy rate by employing semi-definite relaxation and first-order approximation based on Taylor expansion to solve the optimization problem.
Then in order to optimize the sum secrecy rate in the case of the worst estimation error of direction angle toward each eavesdropper, we propose a maximum angle estimation error-SSRM (MAEE-SSRM) method.
The optimization problem is constructed based on the upper and lower bounds of the estimated eavesdropping channel related coefficient and then solved by the change of the variable method.
Simulation results show that our two proposed methods have better sum secrecy rate than zero-forcing (ZF) method and signal-to-leakage-and-noise ratio (SLNR) method.
Furthermore, the sum secrecy rate performance of our VMD-SSRM method is better than that of our MAEE-SSRM method.
\end{abstract}


\IEEEpeerreviewmaketitle

\section{Introduction}
\par
With the explosive growth of the mobile Internet, wireless communications have played an increasingly important role in daily life \cite{H.Zhu, H.Zhu1}.
However, due to the inherent broadcasting nature of the wireless communications, the transmit signals are easily wiretapped by the unauthorized receivers.
Therefore, wireless communications are facing the serious privacy and security problems.
Traditional communication methods based on the cryptographic techniques need an additional secure channel for private key exchanging, which may be insufficient with the development of the mobile Internet \cite{X.Chen, H.Wang2, H.Wang}.
Recently, physical layer security exploiting the characteristics of the wireless channels can realize the secure wireless communications without the upper layer data encryption, being identified as a significant complement to cryptographic techniques \cite{H.Wang1, Y.Zou1, Y.Zou, J.Zhu}.

\par
Directional modulation (DM), which is regarded as one of the promising physical layer wireless security techniques, has attracted wide attentions in recent years \cite{Y.Ding, A.Babakhani, MP.Daly, Y.Ding1, M.Hafez, J.Hu1, F.Shu2, F.Shu3, W.Zhu, B.Gou, Y.Xiao}.
The main ideal of DM is to project the confidential information signals into the desired spatial directions while simultaneously distorting the constellation of the signals in the other directions \cite{Y.Ding}.
In \cite{A.Babakhani}, the authors proposed a DM method which employed the near-field direct antenna modulation technique to overcome the security challenges.
A similar simplified method of DM synthesis was proposed in \cite{MP.Daly} for a far-field scenario.
With the development of the artificial noise (AN) aided technique, DM has been further developed.
The AN is added to the transmit signals expecting that the AN would interfere with the eavesdroppers and not affect the legitimate receivers.
In \cite{Y.Ding1}, a novel baseband signal synthesis method was proposed to construct the AN signal based on the null space of the channel vector in the desired direction.
An orthogonal projection (OP) method in \cite{Y.Ding} was designed for the multi-beam DM systems.
And in \cite{M.Hafez}, the authors proposed a multiple-direction DM transmission scheme which employed the space domain to transmit the multiple data streams independently and increased the capacity of the system.

\par
All the methods mentioned above assumed that the transmitter knows the precise direction angles toward all users (including the eavesdroppers).
However, in practical scenarios, the perfect direction angles are hard to obtain because of the estimation errors induced by the widely-used estimation algorithms such as multiple signal classification (MUSIC) and Capon.
So in \cite{J.Hu} the authors considered a single-user DM system where the estimation error of the direction angle toward the desired receiver was assumed to follow the uniform distribution.
\cite{F.Shu1} considered the imperfect desired direction angles in a multi-beam DM scenario, and assumed the desired direction angle estimation errors follow the truncated Gaussian distribution.

\par
Compared to the direction angles of the desired receivers, the precise direction angles toward the eavesdroppers are more difficult to be obtained as they rarely expose their accurate locations. In this paper, we consider the imperfect direction angles toward eavesdroppers. Moreover, since the Von Mises distribution is regarded as the best statistical model for the direction angles, we assume that the angle estimation error follows the Von Mises distribution \cite{S.Wang}. Meanwhile, the transmit beamforming technique and the AN-aided technique are also employed to realize the secure wireless communications. The main contributions of this paper are summarized as follows:

\par
1) We propose a Von Mises distribution-Sum Secrecy Rate Maximization (VMD-SSRM) robust DM scheme which designs the robust signal beamforming matrix and AN beamforming matrix on the assumption that the estimation errors of direction angles toward the eavesdroppers follow the Von Mises distribution.
First, the expectation of the estimated eavesdropping channel related coefficient is derived. Then the system sum secrecy rate maximization problem subject to the total transmit power of the base station (BS) is formulated. Comprising the logarithms of the product of fractional quadratic functions, the sum secrecy rate expression is non-convex and difficult to tackle. In order to solve this problem, we employ semi-definite relaxation and first-order approximation based on Taylor expansion to transform the original problem into a convex problem.

\par
2) In order to optimize the system sum secrecy rate in the case of the worst estimation error of direction angle toward each eavesdropper, we propose a maximum angle estimation error-SSRM (MAEE-SSRM) method.
To design the robust signal beamforming matrix and AN beamforming matrix, first the upper and lower bounds of the estimated eavesdropping channel related coefficient are derived, then the system sum secrecy rate maximization problem based on the derived upper and lower bounds is constructed. Since the optimization problem is still non-convex, we use the change of the variable method to convert it into a convex problem and then solve it by convex optimization tools.

\par
3) Through simulation and evaluation, we investigate the performance of our two proposed methods. We compare our methods with some typical DM methods such as ZF method \cite{M.Alageli} and SLNR method \cite{F.Shu}. Simulation results show that our two proposed robust methods have much better sum secrecy rate than ZF method and SLNR method. Meanwhile the performance of our VMD-SSRM method is better than that of MAEE-SSRM method.

\par
The rest of this paper is organized as follows. Section II introduces the channel model and the system model.
In Section III, we design the information beamforming matrix and AN beamforming
matrix to maximize the worst-case system sum secrecy rate with the VMD-SSRM method.
In Section IV, we consider the maximum angle estimation error of the eavesdropper and design the information beamforming matrix and AN beamforming matrix to maximize the worst-case system sum secrecy rate with the MAEE-SSRM method. Section V provides the simulation results to validate the effectiveness of the two proposed methods as well as the complexity of our two proposed methods, and followed by our conclusions in Section VI.

\par
\textbf{Notations}: In this paper, we use boldface lowercase and uppercase letters to denote the  vectors and matrices, respectively. $\textbf{A}^{T}$, $\textbf{A}^{H}$, $\mathrm{Tr}(\textbf{A})$, $\mathrm{rank}(\textbf{A})$ and $\mathrm{Re}\{\textbf{A}\}$ denote the transpose, the conjugate transpose, the trace, the rank and the real part of the matrix \textbf{A}, respectively. $\textbf{A}\succeq0$ denotes that $\textbf{A}$ is a positive semi-definite matrix. $|\cdot|$ and $\|{\cdot}\|$ denote the module of the scalars and the Frobenius norm of the matrices. $\mathbb{E}\{\cdot\}$, $\mathrm{log}(\cdot)$, $\mathrm{log}_{10}(\cdot)$ and $\mathrm{In}(\cdot)$ denote the statistical expectation, the logarithm to base 2, the logarithm to base 10 and the natural logarithm, respectively. $\langle{\textbf{x},\textbf{y}}\rangle$ denotes the inner product of the vector \textbf{x} and \textbf{y}. $\mathbb{C}^{{m}\times{n}}$ denotes the set of all complex ${m}\times{n}$ matrices. $\textbf{I}_M$ denotes an ${M}\times{M}$ identity matrix. ${x}\sim{\mathcal{CN}}(0,\sigma^{2})$ denotes a circularly symmetric complex Gaussian random variable $x$ with zero mean and variance $\sigma^{2}$.

\section{system model and problem formulation}

\subsection{Channel Model}

\par
In this paper, the DM transmitter is assumed to be equipped with uniform isotropic array. The array is composed of $N$ elements with an adjacent distance of $l$, while the direction angle between the transmit antenna array and the receiver is denoted by $\theta_{T\!R}$. The normalized steering vector for the transmit antenna array is expressed as
\begin{equation}\label{h_hat_theta_mn}
\hat{\textbf{h}}(\theta_{T\!R})\!=\!\frac{1}{\sqrt{N}}[e^{j2\pi\Psi_{\theta_{\small{T\!R}}}(1)},e^{j2\pi\Psi_{\theta_{T\!R}}(2)},\!\cdots\!,e^{j2\pi\Psi_{\theta_{T\!R}}(N)}]^{T},
\end{equation}
where $\Psi_{\theta_{\small{T\!R}}}(n)$ is
%the phase difference between the $n$-th element and the middle element, and can be denoted as
\begin{equation}\label{Psi_theta_mn}
\Psi_{\theta_{T\!R}}(n)=-\frac{\Big(n-(N+1)/2\Big)l\cos{(\theta_{T\!R})}}{\lambda},n=1,2,\!\cdots\!,N,
\end{equation}
where $\lambda$ is the wavelength of the transmit signal radio.

\par
In the line of sight (LOS) communication scenarios, e.g., unmanned aerial vehicles (UAV) communication and suburban mobile communication, it is widely accepted to use free-space path loss model \cite{A.Al-Hourani}, given by
\begin{equation}\label{PL}
\mathrm{PL}(d)=10\mathrm{log}_{10}\Big(\frac{4\pi df}{c}\Big)^2,
\end{equation}
where $\mathrm{PL}(d)$ is the path loss in dB at a given distance $d$, $f$ is the frequency of the transmit signal radio and $c$ is the radio velocity. Then according to (\ref{h_hat_theta_mn}) and (\ref{PL}), the channel model between the transmit antenna array and the receiver can be expressed as
\begin{equation}\label{h_theta_mn}
\begin{aligned}
&\textbf{h}(\theta_{T\!R})=\sqrt{\Big(\frac{c}{4\pi d_{\small{T\!R}}f}\Big)^2} \times \hat{\textbf{h}}(\theta_{T\!R})\\
&=\sqrt{g_{T\!R}}\frac{1}{\sqrt{N}}[e^{j2\pi\Psi_{\theta_{T\!R}}(1)},e^{j2\pi\Psi_{\theta_{T\!R}}(2)},\cdots,e^{j2\pi\Psi_{\theta_{T\!R}}(N)}]^{T},
\end{aligned}
\end{equation}
where $d_{T\!R}$ is the distance between the transmit antenna array and the receiver, and $g_{T\!R}=\Big(\frac{c}{4\pi d_{\small{T\!R}}f}\Big)^2$ denotes the path loss between the transmit antenna array and the receiver.

\subsection{Multi-User System Model with Multiple Eavesdroppers}
\begin{figure}[!h]
\centering
\includegraphics[width=0.45\textwidth]{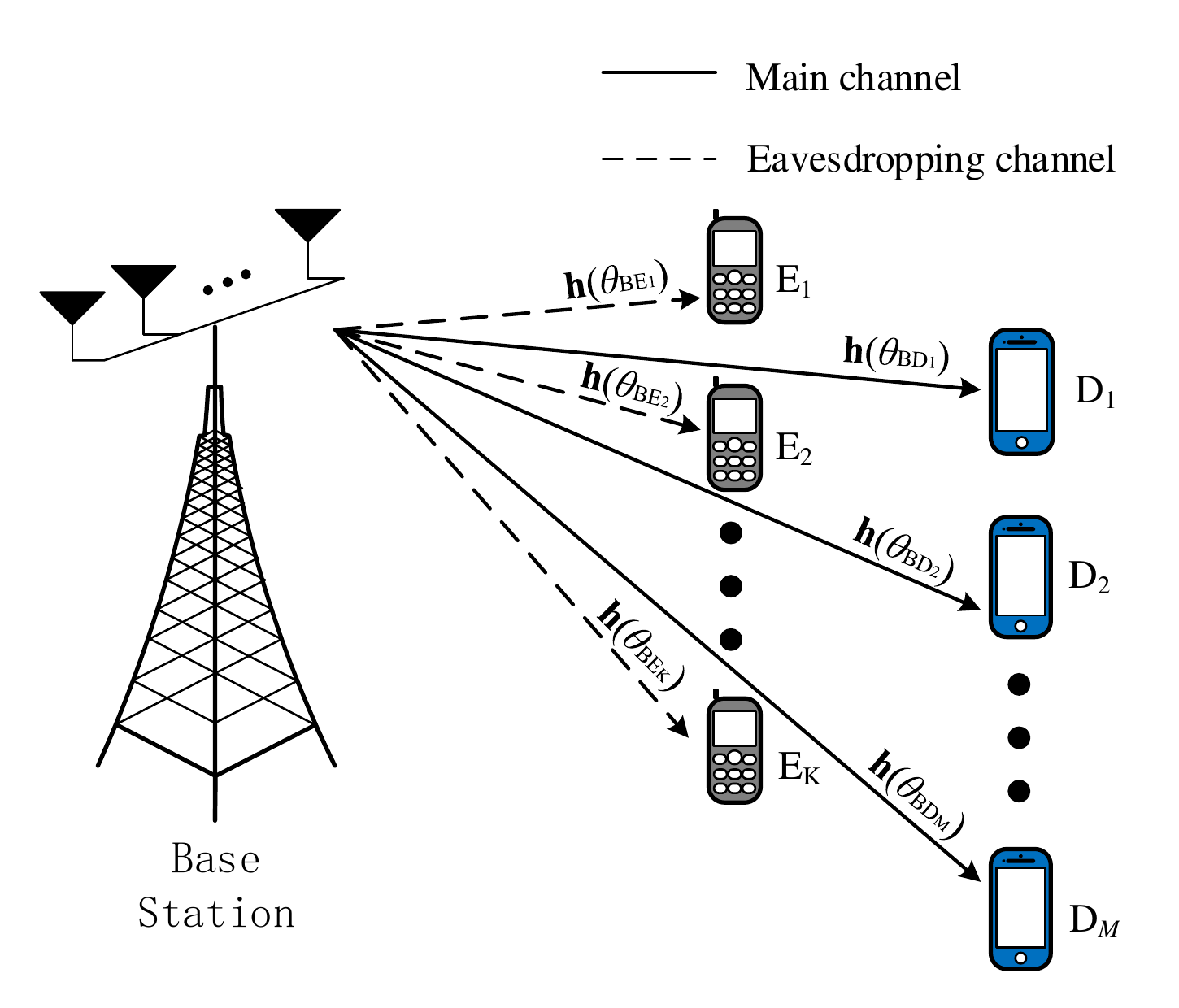}
\caption{Multi-User Downlink System Model with Multiple Eavesdroppers}
\label{fig2}
\end{figure}

\par
In this paper, we consider a flat fading multiuser multi-input single-output (MU-MISO) downlink system as shown in Fig. \ref{fig2}. The system comprises a BS with $N$ transmit antennas, $M$ ($M<N$) single-antenna destination users $D_i$ ($i=1,2,\cdots,M$), and $K$ single-antenna eavesdroppers $E_k$ ($k=1,2,\cdots,K$), wishing to wiretap the confidential information sent from the BS to the destination users. The total transmit power of the BS is $P_t$, and the direction angle between the BS and each destination user is $\theta_{B\!D_i}$ ($i=1,2,\cdots,M$). The transmit beamforming technique is employed by the BS to steer the $M$ information signal beams toward the destination users. The direction angle between the BS and each eavesdropper is $\theta_{B\!E_k}$ ($k=1,2,\cdots,K$). In order to decode the confidential messages sent from the BS to all destination users, the eavesdroppers are assumed to locate closer to the BS than the destination users. And in order to interfere with the eavesdroppers, the BS employs the beamforming technique to steer $L$ ($L\leq{N}$) AN beams toward the eavesdroppers. These AN beams are transmitted along with the information signal beams, thus the received information signal qualities of the eavesdroppers are degraded. Let $\hat{\textbf{x}}=[x_1,\cdots,x_M]^{T}$ denote the signal vector formed by the $M$ independent data streams sent from the BS to the destination users with $\mathbb{E}\{\hat{\textbf{x}}\hat{\textbf{x}}^{H}\}=\textbf{I}_M$. Let $\hat{\textbf{W}}=[\textbf{w}_1,\cdots,\textbf{w}_M]\in\mathbb{C}^{{N}\times{M}}$ and $\hat{\textbf{Q}}=[\textbf{q}_1,\cdots,\textbf{q}_L]\in\mathbb{C}^{{N}\times{L}}$ denote the information beamforming matrix and the AN beamforming matrix, respectively. The total transmit signals of the BS can be expressed as
\begin{equation}\label{x}
\textbf{x}=\sum_{m=1}^{M}\textbf{w}_{m}{x_m}+\sum_{l=1}^{L}\textbf{q}_{l}{z_l},
\end{equation}
where $z_l$ is the AN signal sent by the BS. Then the received signal at the $i$-th destination user $D_i$ is given by
\begin{equation}\label{y_Di}
\begin{aligned}
y_{D_i}&=\textbf{h}^{H}(\theta_{B\!D_i})\textbf{x}+n_{D_i} \\
&=\textbf{h}^{H}(\theta_{B\!D_i})\textbf{w}_{i}{x_i}+\sum\limits_{\substack{m=1\\m\neq{i}}}^{M}\textbf{h}^{H}(\theta_{B\!D_i})\textbf{w}_{m}{x_m} \\
&\quad+\sum_{l=1}^{L}\textbf{h}^{H}(\theta_{B\!D_i})\textbf{q}_{l}{z_l}+n_{D_i},
\end{aligned}
\end{equation}
where the first term of the right hand side of (\ref{y_Di}) is the useful signal received by $D_i$, the second one is the multiuser interference from other destination users, the third one is the AN, and the last one is the additive white Gaussian noise (AWGN) with ${n_{D_i}}\!\sim\!{\mathcal{CN}}(0,\sigma_{D}^{2})$.

\par
Similarly, the signal received at the $k$-th eavesdropper $E_k$ is
\begin{equation}\label{y_Ek}
\begin{aligned}
y_{E_k}&=\textbf{h}^{H}(\theta_{B\!E_k})\textbf{x}+n_{E_k} \\
&=\textbf{h}^{H}(\theta_{B\!E_k})\textbf{w}_{i}{x_i}+\sum\limits_{\substack{m=1\\m\neq{i}}}^{M}\textbf{h}^{H}(\theta_{B\!E_k})\textbf{w}_{m}{x_m}\\
&\quad+\sum_{l=1}^{L}\textbf{h}^{H}(\theta_{B\!E_k})\textbf{q}_{l}{z_l}+n_{E_k},
\end{aligned}
\end{equation}
where the right hand side of (\ref{y_Ek}) is composed of the $i$-th confidential message intercepted by $E_k$, the messages sent to the other ($M$-1) destination users that are intercepted by $E_k$, the AN to disturb $E_k$ and the AWGN satisfying ${n_{E_k}}\!\sim\!{\mathcal{CN}}(0,\sigma_{E}^{2})$.
\par
It should be noted that we consider the non-colluding eavesdropping scenario where each eavesdropper relies on itself to overhear and decode the messages and there is no information exchange between the eavesdroppers.

\par
In this paper, the worst-case system sum secrecy rate is used to evaluate the system performance in the multiuser scenario, which can be expressed as
\begin{equation}\label{R_s}
\begin{aligned}
R_s&=\sum_{i=1}^{M}\underset{k}{\mathop{\min}}~R_{i}^k  \\
&=\sum_{i=1}^{M}\underset{k}{\mathop{\min}}~\Big(\!\log(1+\mathrm{SINR}_{D_i})\!-\!\log(1+\mathrm{SINR}_{E_k})\!\Big) \\
&=\!\log\!\Big(\!\prod_{i=1}^{M}(1+\mathrm{SINR}_{D_i})\!\Big)\!-\!\log\!\Big(\!\prod_{i=1}^{M}\underset{k}{\mathop{\max}} (1+\mathrm{SINR}_{E_k})\!\Big),
\end{aligned}
\end{equation}
where $R_s$ denotes the system sum secrecy rate, $\mathrm{SINR}_{D_i}$ and $\mathrm{SINR}_{E_k}$  denote the signal-to-interference-plus-noise ratio (SINR) at the destination user $D_i$ and the eavesdropper $E_k$, respectively.

\par
Then, according to (\ref{y_Di}), the SINR at the $i$-th destination user $D_i$ is
\begin{equation}\label{SINR_Di}
\mathrm{SINR}_{D_i}\!=\!\frac{|\textbf{h}^{H}(\theta_{B\!D_i})\textbf{w}_{i}|^{2}}{\sum\limits_{\substack{m=1\\m\neq{i}}}^{M}|\textbf{h}^{H}(\theta_{B\!D_i})\textbf{w}_{m}|^{2}   \!+\!\sum\limits_{l=1}^{L}|\textbf{h}^{H}(\theta_{B\!D_i})\textbf{q}_{l}|^{2}\!+\!\sigma_D^{2}}.
\end{equation}
\par
And according to (\ref{y_Ek}), the SINR at the $k$-th eavesdropper $E_k$ intending to intercept the $i$-th destination user is
\begin{equation}\label{SINR_Ek}
\mathrm{SINR}_{E_k}\!=\!\frac{|\textbf{h}^{H}(\theta_{B\!E_k})\textbf{w}_{i}|^{2}}{\sum\limits_{\substack{m=1\\m\neq{i}}}^{M}|\textbf{h}^{H}(\theta_{B\!E_k})\textbf{w}_{m}|^{2}   \!+\!\sum\limits_{l=1}^{L}|\textbf{h}^{H}(\theta_{B\!E_k})\textbf{q}_{l}|^{2}\!+\!\sigma_E^{2}}.
\end{equation}

\par
Now, according to (\ref{R_s}), (\ref{SINR_Di}) and (\ref{SINR_Ek}), the optimization problem that maximizes the worst-case system sum secrecy rate under the constraint of the total transmit power of the BS is constructed as shown in (\ref{SSRM}) on the following page.

\newcounter{mytempeqncnt2}
\begin{figure*}[!t]
\normalsize
\setcounter{equation}{10}
\begin{align}\label{SSRM}
&\underset{\mathbf{\mathit{\{\textbf{w}_{i}\},\{\textbf{q}_{l}\}}}}{\mathop{\mathrm{maximize}}} \nonumber \\
&\log\!\Big(\!\prod_{i=1}^{M}\frac{\sum\limits_{m=1}^{M}|\textbf{h}^{H}(\theta_{B\!D_i})\textbf{w}_{m}|^{2}\!+\!\sum\limits_{l=1}^{L}|\textbf{h}^{H}(\theta_{B\!D_i})\textbf{q}_{l}|^{2}\!+\!\sigma_{D}^{2}}
{\sum\limits_{\substack{m=1\\m\neq{i}}}^{M}|\textbf{h}^{H}(\theta_{B\!D_i})\textbf{w}_{m}|^{2}\!+\!\sum\limits_{l=1}^{L}|\textbf{h}^{H}(\theta_{B\!D_i})\textbf{q}_{l}|^{2}\!+\!\sigma_{D}^{2}}\Big)\!-
\log\!\Big(\!\prod_{i=1}^{M}\!\underset{k}{\mathop{\max}}\!\frac{\sum\limits_{m=1}^{M}|\textbf{h}^{H}(\theta_{B\!E_k})\textbf{w}_{m}|^{2}\!+\!\sum\limits_{l=1}^{L}|\textbf{h}^{H}(\theta_{B\!E_k})\textbf{q}_{l}|^{2}\!+\!\sigma_{E}^{2}}
{\sum\limits_{\substack{m=1\\m\neq{i}}}^{M}|\textbf{h}^{H}(\theta_{B\!E_k})\textbf{w}_{m}|^{2}\!+\!\sum\limits_{l=1}^{L}|\textbf{h}^{H}(\theta_{B\!E_k})\textbf{q}_{l}|^{2}\!+\!\sigma_{E}^{2}}\!\Big)
\nonumber \\
&\mathrm{s.t.}~\sum\limits_{l=1}^{L}\|\textbf{q}_{l}\|^{2}+\sum\limits_{i=1}^{M}\|\textbf{w}_{i}\|^{2}\leq{P_t}.
\end{align}
\hrulefill
\vspace*{4pt}
\end{figure*}

\subsection{Direction Angle Estimation Error Following Von Mises Distribution}
\par
Since the BS is assumed to have the imperfect knowledge of the direction angles toward the eavesdroppers, the actual direction angle between the BS and the $k$-th eavesdropper is expressed as
\setcounter{equation}{11}
\begin{equation}\label{theta_BEk}
\theta_{B\!E_k}=\hat{\theta}_{B\!E_k}+\Delta\theta_{B\!E_k},k=1,\cdots,K,
\end{equation}
where $\hat{\theta}_{B\!E_k}$ is the estimated direction angle between the BS and the $k$-th eavesdropper, $\Delta\theta_{B\!E_k}$ denotes the estimation error which is assumed to follow the Von Mises distribution \cite{KV.Mardia} over the interval $[-\Delta\theta_{\mathrm{max}},\Delta\theta_{\mathrm{max}}]$. Note that $\Delta\theta_{\mathrm{max}}$ represents the maximum angle error and is a positive value up to the beamwidth between the first nulls \cite{J.Hu}.

The Von Mises distribution, also known as the circular normal distribution, is a continuous probability distribution on the circle and mainly describes the information related to the directional angles. In \cite{S.Wang}, the Von Mises distribution was considered to be the best statistical model for circular parameters such as the direction angle. Subsequently, according to (\ref{theta_BEk}), the estimation error of the direction angle should also follow the Von Mises distribution.

\par
The probability density function (PDF) of the Von Mises distribution is
%{
%\setlength\abovedisplayskip{1pt}
%\setlength\belowdisplayskip{1pt}
\begin{equation}\label{f_theta}
f(\theta|\mu,\kappa)=\frac{e^{\kappa\cos(\theta-\mu)}}{2\pi{I_0}(\kappa)},\theta\in[-\pi,\pi),
\end{equation}
%}%\\表示换行，两个&表示公式间对齐的位置，\begin{equation}...\end{equation} 自动给公式编号。
where $\theta$ represents the estimation error of the direction angle, $\mu$ denotes the mean of $\theta$, ${\kappa}\geq{0}$ denotes the concentration parameter and controls the width of $\theta$, and $I_n$ with $n=0$ is the modified Bessel function of the first kind with order $n$ computed by
\begin{equation}\label{In_x}
I_n(x)=\frac{1}{\pi}\int_{0}^{\pi}e^{x\cos(t)}\cos(nt)dt,x\in(-\infty,\infty).
\end{equation}

\section{ROBUST BEAMFORMING MATRIX DESIGN WITH VON-MISES DISTRIBUTED DIRECTION ANGLE ERROR AT EAVESDROPPERS}

\par
In the previous section, the estimation error of the direction angle toward each eavesdropper has been modeled to follow the Von Mises distribution. Then in order to maximize the worst-case system sum secrecy rate, in this section, we propose a Von Mises distribution-SSRM (VMD-SSRM) robust DM scheme. The procedure of the proposed scheme, i.e., the design of the optimal beamforming matrices, will be illustrated as follows.

\par
In (\ref{SSRM}), the item $|\textbf{h}^{H}(\theta_{B\!E_k})\textbf{w}_{m}|^{2}$ can be expressed as
\begin{equation}
\begin{aligned}
&|\textbf{h}^{H}(\theta_{B\!E_k})\textbf{w}_{m}|^{2}=\\
&\mathrm{Tr}\Big(\textbf{h}(\theta_{B\!E_k})\textbf{h}^{H}(\theta_{B\!E_k})\textbf{w}_{m}\textbf{w}_{m}^{H}\Big)\\
&=\mathrm{Tr}\Big(\textbf{h}(\hat{\theta}_{B\!E_k}+\Delta\theta_{B\!E_k})\textbf{h}^{H}(\hat{\theta}_{B\!E_k}+\Delta\theta_{B\!E_k})\textbf{w}_{m}\textbf{w}_{m}^{H}\Big).
\end{aligned}
\end{equation}

\par
To simplify the expression, we define
\begin{equation}\label{H_BEk}
\textbf{H}_{B\!E_k}\!=\!\mathbb{E}\Big\{\textbf{h}(\hat{\theta}_{B\!E_k}+\Delta\theta_{B\!E_k}){\textbf{h}^{H}(\hat{\theta}_{B\!E_k}+\Delta\theta_{B\!E_k})}\Big\},k\in{K},   \end{equation}
where $\textbf{H}_{B\!E_k}\in{\mathbb{C}^{{N}\times{N}}}$. The necessity of the expectation in (\ref{H_BEk}) is explained as follows. If we remove the expectation, since $\Delta\theta_{B\!E_k}$ is a random variable, then $\textbf{H}_{B\!E_k}$ would be random as well as the objective function of (\ref{SSRM}), and it would be very difficult to find a solution to (\ref{SSRM}).

\par
Let $H_{B\!E_k}(u,v)$ denote the $u$-th row and the $v$-th column entry of $\textbf{H}_{B\!E_k}$, then $H_{B\!E_k}(u,v)$ can be written as
\begin{equation}\label{H_BEk_pq}
H_{B\!E_k}(u,v)=\Upsilon_{1k}(u,v)-j\Upsilon_{2k}(u,v).
\end{equation}
The detailed procedure for deriving $\Upsilon_{1k}(u,v)$ and $\Upsilon_{2k}(u,v)$ based on the Von Mises distribution is shown in Appendix A.
And $\Upsilon_{1k}(u,v)$ and $\Upsilon_{2k}(u,v)$ are expressed as (\ref{Gamma_1k}) and (\ref{Gamma_2k}), respectively.

\par
In order to make (\ref{SSRM}) more tractable, we substitute the new matrix $\textbf{H}(\theta_{B\!D_i})=\textbf{h}(\theta_{B\!D_i})\textbf{h}^{H}(\theta_{B\!D_i})$ as well as the positive semi-definite matrix variables $\textbf{W}_i=\textbf{w}_i\textbf{w}^{H}_i$, $\textbf{Q}=\sum\limits_{l=1}^{L}\textbf{q}_l\textbf{q}^{H}_l$ into (\ref{SSRM}). Then the optimization problem (\ref{SSRM}) can be rewritten in a simple form as shown in (\ref{SSRM_1}) at the top of the next page.
\newcounter{mytempeqncnt3}
\begin{figure*}[!t]
\normalsize
\setcounter{equation}{17}
\begin{align}\label{SSRM_1}
&\underset{\mathbf{\mathit{\{\textbf{W}_{i}\},\textbf{Q}}}}{\mathop{\mathrm{maximize}}}
\nonumber \\
&\log\!\Big(\!\prod_{i=1}^{M}\!\frac{\sum\limits_{m=1}^{M}\!\mathrm{Tr}\!\Big(\textbf{H}(\theta_{B\!D_i})\textbf{W}_m\Big)\!+\!\mathrm{Tr}\!\Big(\textbf{H}(\theta_{B\!D_i})\textbf{Q}\Big)\!+\!\sigma_{D}^{2}}
{\sum\limits_{\substack{m=1\\m\neq{i}}}^{M}\!\mathrm{Tr}\!\Big(\textbf{H}(\theta_{B\!D_i})\textbf{W}_m\Big)\!+\!\mathrm{Tr}\!\Big(\textbf{H}(\theta_{B\!D_i})\textbf{Q}\Big)\!+\!\sigma_{D}^{2}}\Big)\!-
\log\!\Big(\!\prod_{i=1}^{M}\underset{k}{\mathop{\max}}\frac{\sum\limits_{m=1}^{M}\!\mathrm{Tr}\Big(\textbf{H}_{B\!E_k}\textbf{W}_m\Big)\!+\!\mathrm{Tr}\Big(\textbf{H}_{B\!E_k}\textbf{Q}\Big)\!+\!\sigma_{E}^{2}}
{\sum\limits_{\substack{m=1\\m\neq{i}}}^{M}\!\mathrm{Tr}\Big(\textbf{H}_{B\!E_k}\textbf{W}_m\Big)\!+\!\mathrm{Tr}\Big(\textbf{H}_{B\!E_k}\textbf{Q}\Big)\!+\!\sigma_{E}^{2}}\Big)
\nonumber \\
&\mathrm{s.t.}~\mathrm{Tr}(\textbf{Q})+\sum\limits_{i=1}^{M}\mathrm{Tr}(\textbf{W}_i)\leq{P_t},
\nonumber \\
&{\phantom{=}\;\;}~\textbf{W}_i,\textbf{Q}\succeq\textbf{0},\forall{i}.
\end{align}
\hrulefill
\vspace*{4pt}
\end{figure*}

\par
By solving the problem (\ref{SSRM_1}), we can obtain the optimal information beamforming matrices $\{\textbf{W}_i\}$ and the AN beamforming matrix \textbf{Q}.
%The information beamforming matrices $\{\textbf{W}_i\}$ should aim to solve a contradicting balance between the following: 1) increasing the information signal power at the destination users; 2) cancelling the inter-user interference at the destination users; 3) reducing the information signal leakage to the eavesdroppers. On the other hand, the AN beamforming matrix $\textbf{Q}$ should be optimized to jam the eavesdroppers. Since the power budget at BS is limited, the power allocation between $\Big(\sum\limits_{i=1}^{M}\mathrm{Tr}(\textbf{W}_i)\Big)$ and $\Big(\mathrm{Tr}(\textbf{Q})\Big)$ should be held in the optimal balance, thus the constraint condition in (\ref{SSRM_1}) can both satisfy the power constraints and ensure the maximum sum secrecy rate of the system.
However, the objective function in (\ref{SSRM_1}) is non-convex and difficult to tackle, as it comprises the logarithms of the product of fractional quadratic functions. In order to solve this problem, we employ the semi-definite relaxation and the first-order approximation technique based on the Taylor expansion to transform the original problem into a convex problem \cite{P.Zhao}. Then this convex problem can be solved by using the convex optimization toolbox (CVX).

\par
The specific process for converting (\ref{SSRM_1}) into a convex one is illustrated as follows. First, the exponential variables are used to substitute the numerators and denominators of the fractions in the objective function in (\ref{SSRM_1}), i.e.,
\begin{equation}\label{exp_pi}
e^{p_i}=\sum\limits_{m=1}^{M}\mathrm{Tr}\Big(\textbf{H}(\theta_{B\!D_i})\textbf{W}_m\Big)+\mathrm{Tr}\Big(\textbf{H}(\theta_{B\!D_i})\textbf{Q}\Big)+\sigma_{D}^{2},
\end{equation}
\begin{equation}\label{exp_qi}
e^{q_i}=\sum\limits_{\substack{m=1\\m\neq{i}}}^{M}\mathrm{Tr}\Big(\textbf{H}(\theta_{B\!D_i})\textbf{W}_m\Big)+\mathrm{Tr}\Big(\textbf{H}(\theta_{B\!D_i})\textbf{Q}\Big)+\sigma_{D}^{2},
\end{equation}
\begin{equation}\label{exp_ck}
e^{c_k}=\sum\limits_{m=1}^{M}\mathrm{Tr}\Big(\textbf{H}_{B\!E_k}\textbf{W}_m\Big)+\mathrm{Tr}\Big(\textbf{H}_{B\!E_k}\textbf{Q}\Big)+\sigma_{E}^{2},
\end{equation}
\begin{equation}\label{exp_dik}
e^{d_{i,k}}=\sum\limits_{\substack{m=1\\m\neq{i}}}^{M}\mathrm{Tr}\Big(\textbf{H}_{B\!E_k}\textbf{W}_m\Big)+\mathrm{Tr}\Big(\textbf{H}_{B\!E_k}\textbf{Q}\Big)+\sigma_{E}^{2}.
\end{equation}

\par
Then according to the properties of exponential and logarithmic functions, the objective function in (\ref{SSRM_1}) can be rewritten as
\begin{equation}\label{Obj_fun}
\sum\limits_{i=1}^{M}\big((p_i-q_i)-\underset{k=1,\cdots,K}{\mathop{\max}}(c_k-d_{i,k})\big).
\end{equation}
Meanwhile $p_i$, $q_i$, $c_k$, $d_{i,k}$ are constrained by the right hand sides of each equation in (\ref{exp_pi}), (\ref{exp_qi}), (\ref{exp_ck}) and (\ref{exp_dik}). So the problem (\ref{SSRM_1}) can be transformed to a semi-definite programming (SDP) problem as

%Therefore, by defining the following real-valued slack variables
%\begin{subequations}
%\begin{equation}
%\textbf{x}=[x_1,\cdots,x_M]^{T},\textbf{y}=[y_1,\cdots,y_M]^{T},
%\end{equation}
%\begin{equation}
%\textbf{u}=[u_1,\cdots,u_K]^{T},\textbf{V}=\left(\begin{array}{ccc}
%v_{1,1} & \ldots & v_{1,K}\\
%\vdots & \ddots & \vdots \\
%v_{M,1} & \ldots & v_{M,K}
%\end{array}\right),
%\end{equation}
%\end{subequations}
%and the set of optimization variables $\mathbb{S}=\Big\{\{\textbf{W}_m\},\textbf{Q},\textbf{x},\textbf{y},\textbf{u},\\
%\textbf{V}\Big\},m=1,\cdots,M$, the semi-definite programming (SDP) reformulation of the problem (\ref{SSRM_1}) is
\begin{subequations}\label{SSRM_2}
\begin{align}
&\underset{\mathbb{S}}{\mathop{\mathrm{maximize}}}~ \sum\limits_{i=1}^{M}\big((p_i-q_i)-\underset{k=1,\cdots,K}{\mathop{\max}}(c_k-d_{i,k})\big)\\
&\mathrm{s.t.} \nonumber \\
&\sum\limits_{m=1}^{M}\mathrm{Tr}\Big(\textbf{H}(\theta_{B\!D_i})\textbf{W}_m\Big)+\mathrm{Tr}\Big(\textbf{H}(\theta_{B\!D_i})\textbf{Q}\Big)+\sigma_{D}^{2}\geq{e^{p_i}},\forall{i},\\
&\sum\limits_{\substack{m=1\\m\neq{i}}}^{M}\mathrm{Tr}\Big(\textbf{H}(\theta_{B\!D_i})\textbf{W}_m\Big)+\mathrm{Tr}\Big(\textbf{H}(\theta_{B\!D_i})\textbf{Q}\Big)+\sigma_{D}^{2}\leq{e^{q_i}},\forall{i},\\
&\sum\limits_{m=1}^{M}\mathrm{Tr}\Big(\textbf{H}_{B\!E_k}\textbf{W}_m\Big)+\mathrm{Tr}\Big(\textbf{H}_{B\!E_k}\textbf{Q}\Big)+\sigma_{E}^{2}\leq{e^{c_k}},\forall{k},\\
&\sum\limits_{\substack{m=1\\m\neq{i}}}^{M}\mathrm{Tr}\Big(\textbf{H}_{B\!E_k}\textbf{W}_m\Big)+\mathrm{Tr}\Big(\textbf{H}_{B\!E_k}\textbf{Q}\Big)+\sigma_{E}^{2}\geq{e^{d_{i,k}}},\forall{i},\forall{k},\\
&\mathrm{Tr}(\textbf{Q})+\sum\limits_{i=1}^{M}\mathrm{Tr}(\textbf{W}_i)\leq{P_t},\\
&\textbf{W}_i,\textbf{Q}\succeq\textbf{0},\forall{i}.
\end{align}
\end{subequations}

\par
The objective function (\ref{SSRM_2}a) is a concave function because it comprises a sum of the affine functions minus a sum of the maxima of the affine functions. However, the constraints (\ref{SSRM_2}c) and (\ref{SSRM_2}d) are non-convex. In order to transform these two constraints into convex ones, $e^{q_i}$ and $e^{c_k}$ can be linearized based on the first-order Taylor approximation as
\begin{equation}
e^{q_i}=e^{\bar{q}_i}(q_i-\bar{q}_i+1),
\end{equation}
\begin{equation}
e^{c_k}=e^{\bar{c}_k}(c_k-\bar{c}_k+1),
\end{equation}
where
\begin{equation}\label{bar_y}
\bar{q}_i=\mathrm{ln}\Big(\sum\limits_{\substack{m=1\\m\neq{i}}}^{M}\mathrm{Tr}\Big(\textbf{H}(\theta_{B\!D_i})\textbf{W}_m\Big)\!+\!\mathrm{Tr}\Big(\textbf{H}(\theta_{B\!D_i})\textbf{Q}\Big)\!+\!\sigma_{D}^{2}\Big),
\end{equation}
\begin{equation}\label{bar_u}
\bar{c}_k=\mathrm{ln}\Big(\sum\limits_{m=1}^{M}\mathrm{Tr}\Big(\textbf{H}_{B\!E_k}\textbf{W}_m\Big)+\mathrm{Tr}\Big(\textbf{H}_{B\!E_k}\textbf{Q}\Big)+\sigma_{E}^{2}\Big),
\end{equation}
are the  points around which the approximation are made. So the problem (\ref{SSRM}) eventually becomes
\begin{subequations}\label{SSRM_3}
\begin{align}
&\underset{\mathbb{S}}{\mathop{\mathrm{maximize}}}~ \sum\limits_{i=1}^{M}\big((p_i-q_i)-\underset{k=1,\cdots,K}{\mathop{\max}}(c_k-d_{i,k})\big)\\
&\mathrm{s.t.} \nonumber \\
&\sum\limits_{m=1}^{M}\mathrm{Tr}\Big(\textbf{H}(\theta_{BD_i})\textbf{W}_m\Big)+\mathrm{Tr}\Big(\textbf{H}(\theta_{BD_i})\textbf{Q}\Big)+\sigma_{D}^{2}\geq{e^{p_i}},\forall{i},\\
&\sum\limits_{\substack{m=1\\m\neq{i}}}^{M}\mathrm{Tr}\Big(\textbf{H}(\theta_{BD_i})\textbf{W}_m\Big)+\mathrm{Tr}\Big(\textbf{H}(\theta_{BD_i})\textbf{Q}\Big)+\sigma_{D}^{2} \nonumber \\
&\leq{e^{\bar{q}_i}(q_i-\bar{q}_i+1)},\forall{i},\\
&\sum\limits_{m=1}^{M}\mathrm{Tr}\Big(\textbf{H}_{B\!E_k}\textbf{W}_m\Big)+\mathrm{Tr}\Big(\textbf{H}_{B\!E_k}\textbf{Q}\Big)+\sigma_{E}^{2} \nonumber \\
&\leq{e^{\bar{c}_k}(c_k-\bar{c}_k+1),\forall{k},}\\
&\sum\limits_{\substack{m=1\\m\neq{i}}}^{M}\mathrm{Tr}\Big(\textbf{H}_{B\!E_k}\textbf{W}_m\Big)+\mathrm{Tr}\Big(\textbf{H}_{B\!E_k}\textbf{Q}\Big)+\sigma_{E}^{2}\geq{e^{d_{i,k}}},\forall{i},\forall{k},\\
&\mathrm{Tr}(\textbf{Q})+\sum\limits_{i=1}^{M}\mathrm{Tr}(\textbf{W}_i)\leq{P_t},\\
&\textbf{W}_i,\textbf{Q}\succeq\textbf{0},\forall{i}.
\end{align}
\end{subequations}

\par
As a convex problem, (\ref{SSRM_3}) can be solved iteratively by the Algorithm 1 using the CVX optimization software. According to the definition, $\textbf{W}_m$ has to satisfy $\mathrm{rank}(\textbf{W}_m)=1$, so the optimal $\textbf{W}_m$ of (\ref{SSRM_3}) should satisfy this rank-one constraint. Otherwise the randomization technology \cite{M.Zhang} would be utilized to get a rank-one approximation.
\begin{table}[h]
\centering
\begin{tabular}{l}
\hline
\textbf{Algorithm 1} \quad Algorithm for solving the problem (\ref{SSRM_3}) \\
\hline
1:  Given $\{\tilde{\textbf{w}}_i,i=1,\cdots,M\}$ and $\{\tilde{\textbf{q}}_l,l=1,\cdots,L\}$ randomly that\\
\quad are feasible to (\ref{SSRM_3}); \\
2: Set $\tilde{\textbf{W}}_i[0]=\tilde{\textbf{w}}_{i}\tilde{\textbf{w}}_{i}^{H},i=1,\cdots,M$, $\tilde{\textbf{Q}}[0]=\sum\limits_{l=1}^{L}\tilde{\textbf{q}}_{l}\tilde{\textbf{q}}_{l}^{H}$ and set $n$=0; \\
3:  \textbf{Repeat}\\
4:  Substituting $\tilde{\textbf{W}}_i[n]$ and $\tilde{\textbf{Q}}[n]$ into (\ref{bar_y}) and (\ref{bar_u}) yields $\bar{q}_i[n+1]$ \\ \quad and $\bar{c}_k[n+1]$;  \\
5:  Increment $n=n+1$;  \\
6:  Substituting $\bar{q}_i[n]$ and $\bar{c}_k[n]$ into (\ref{SSRM_3}) yields the optimal solution \\ \quad $\tilde{\textbf{W}}_i[n]$ and $\tilde{\textbf{Q}}[n]$;  \\
7:  \textbf{Until} convergence; \\
8:  Obtain $\tilde{\textbf{Q}}$ and $\{\textbf{w}_{i}^{*},i=1,\cdots,M\}$ by decomposition of \\
\quad $\tilde{\textbf{W}}_i[n]=(\textbf{w}_{i}^{*})(\textbf{w}_{i}^{*})^{H}$ for all $i$ in the case of $\mathrm{rank}(\tilde{\textbf{W}}_i[n])=1$; \\
\quad otherwise the randomization technology \cite{M.Zhang} would be utilized to get \\ \quad a rank-one approximation.\\
\hline
\end{tabular}
\end{table}

\section{ROBUST BEAMFORMING MATRIX DESIGN WITH MAXIMUM DIRECTION ANGLE ERROR AT EAVESDROPPERS}
\par
In the previous section, the system was designed based on the Von Mises distribution of the eavesdropper direction angle error. Thus the system sum secrecy rate was maximized for all possible values of the direction angle error. Nevertheless, it is also important to exclusively investigate the case of the worst direction angle error. Thus in this section, the system will be designed when the eavesdropper direction angle error reaches its maximum value and the robust beamforming matrices are obtained by the maximum angle estimation error-SSRM (MAEE-SSRM) method.

\par
If the direction angle estimation error toward each eavesdropper is bounded, we can prove that the channel estimation error between the BS and each eavesdropper is norm-bounded (the detailed procedure is illustrated in Appendix B). Then the channel between the BS and the $k$-th eavesdropper can be expressed as
\begin{equation}
\textbf{h}(\theta_{B\!E_k})=\textbf{h}(\hat{\theta}_{B\!E_k})+\Delta\textbf{h}_{B\!E_k},\|\Delta\textbf{h}_{B\!E_k}\|\leq\varepsilon_k,k=1,\!\cdots\!,K,
\end{equation}
where $\textbf{h}(\hat{\theta}_{B\!E_k})$ is the estimated channel between the BS and the $k$-th eavesdropper, $\Delta\textbf{h}_{B\!E_k}$ is the channel estimation error, and $\varepsilon_k$ is the bound of the norm of the channel estimation error of the $k$-th eavesdropper.

\par
Then we derive the lower bound of the objective function in (\ref{SSRM}). For that purpose, we first derive the upper and lower bounds of $|\textbf{h}^{H}(\theta_{B\!E_k})\textbf{w}_m|^{2}$ and $|\textbf{h}^{H}(\theta_{B\!E_k})\textbf{q}_l|^{2}$ in (\ref{SSRM}). The upper bound of $|\textbf{h}^{H}(\theta_{B\!E_k})\textbf{w}_m|^{2}$ is
\begin{equation}\label{Upper_B}
\begin{aligned}
&|\textbf{h}^{H}(\theta_{B\!E_k})\textbf{w}_m|^{2} \\
&=\textbf{h}^{H}(\theta_{B\!E_k})\textbf{w}_m\textbf{w}_m^{H}\textbf{h}(\theta_{B\!E_k})\\
&=\Big(\textbf{h}^{H}(\hat{\theta}_{B\!E_k})+\Delta\textbf{h}_{B\!E_k}^{H}\Big)\textbf{w}_m\textbf{w}_m^{H}\Big(\textbf{h}(\hat{\theta}_{B\!E_k})+\Delta\textbf{h}_{B\!E_k}\Big)\\
&=\textbf{h}^{H}(\hat{\theta}_{B\!E_k})\textbf{w}_m\textbf{w}_m^{H}\textbf{h}(\hat{\theta}_{B\!E_k})\!+\!2\mathrm{Re}\!\Big\{\!\Delta\textbf{h}_{B\!E_k}^{H}\textbf{w}_m\textbf{w}_m^{H}\textbf{h}(\hat{\theta}_{B\!E_k})\!\Big\}\\
&\overset{(a)}{\leq}\textbf{h}^{H}(\hat{\theta}_{B\!E_k})\textbf{w}_m\textbf{w}_m^{H}\textbf{h}(\hat{\theta}_{B\!E_k})+2\varepsilon_k\|\textbf{w}_m\textbf{w}_m^{H}\textbf{h}(\hat{\theta}_{B\!E_k})\|.\\
\end{aligned}
\end{equation}
\noindent Similarly, the lower bound of $|\textbf{h}^{H}(\theta_{B\!E_k})\textbf{w}_m|^{2}$ is
\begin{equation}\label{Lower_B}
\begin{aligned}
&|\textbf{h}^{H}(\theta_{B\!E_k})\textbf{w}_m|^{2} \\
&=\textbf{h}^{H}(\hat{\theta}_{B\!E_k})\textbf{w}_m\textbf{w}_m^{H}\textbf{h}(\hat{\theta}_{B\!E_k})\!+\!2\mathrm{Re}\!\Big\{\!\Delta\textbf{h}_{B\!E_k}^{H}\textbf{w}_m\textbf{w}_m^{H}\textbf{h}(\hat{\theta}_{B\!E_k})\!\Big\}\\
&\geq\textbf{h}^{H}(\hat{\theta}_{B\!E_k})\textbf{w}_m\textbf{w}_m^{H}\textbf{h}(\hat{\theta}_{B\!E_k})-2\varepsilon_k\|\textbf{w}_m\textbf{w}_m^{H}\textbf{h}(\hat{\theta}_{B\!E_k})\|.\\
\end{aligned}
\end{equation}

\par
It should be noted that in (\ref{Upper_B}) and (\ref{Lower_B}) the second-order error term $\Delta\textbf{h}_{B\!E_k}^{H}\textbf{w}_m\textbf{w}_m^{H}\Delta\textbf{h}_{B\!E_k}$ is omitted because it is quite small compared to the other two terms. And note that the step ($a$) in (\ref{Upper_B}) results from the following derivation process. According to the Cauchy-Schwarz inequality theorem, we have
\begin{equation}\label{CS_2}
|\textbf{x}^{H}\textbf{y}|\leq\|\textbf{x}\|\|\textbf{y}\|.
\end{equation}
Since the inequality
\begin{equation}\label{CS_3}
-|\textbf{x}^{H}\textbf{y}|\leq\mathrm{Re}\{\textbf{x}^{H}\textbf{y}\}\leq|\textbf{x}^{H}\textbf{y}|,
\end{equation}
is obviously true, then we can obtain the following inequality
\begin{equation}\label{CS_4}
-\|\textbf{x}\|\|\textbf{y}\|\leq\mathrm{Re}\{\textbf{x}^{H}\textbf{y}\}\leq\|\textbf{x}\|\|\textbf{y}\|.
\end{equation}
\noindent The inequality in (\ref{CS_4}) holds with equality when and only when $\textbf{x}$ and $\textbf{y}$ are linearly dependent. The derivation of the step ($a$) in (\ref{Upper_B}) is completed.

\par
Similarly, the upper bound of $|\textbf{h}^{H}(\theta_{B\!E_k})\textbf{q}_l|^{2}$ is
\begin{equation}
\begin{aligned}
&|\textbf{h}^{H}(\theta_{B\!E_k})\textbf{q}_l|^{2} \\
&\leq\textbf{h}^{H}(\hat{\theta}_{B\!E_k})\textbf{q}_l\textbf{q}_l^{H}\textbf{h}(\hat{\theta}_{B\!E_k})+2\varepsilon_k\|\textbf{q}_l\textbf{q}_l^{H}\textbf{h}(\hat{\theta}_{B\!E_k})\|,\\
\end{aligned}
\end{equation}
and the lower bounds of $|\textbf{h}^{H}(\theta_{B\!E_k})\textbf{q}_l|^{2}$ is
\begin{equation}
\begin{aligned}
&|\textbf{h}^{H}(\theta_{B\!E_k})\textbf{q}_l|^{2} \\
&\geq\textbf{h}^{H}(\hat{\theta}_{B\!E_k})\textbf{q}_l\textbf{q}_l^{H}\textbf{h}(\hat{\theta}_{B\!E_k})-2\varepsilon_k\|\textbf{q}_l\textbf{q}_l^{H}\textbf{h}(\hat{\theta}_{B\!E_k})\|.\\
\end{aligned}
\end{equation}

\par
With the upper and lower bounds of $|\textbf{h}^{H}(\theta_{B\!E_k})\textbf{w}_m|^{2}$ and $|\textbf{h}^{H}(\theta_{B\!E_k})\textbf{q}_l|^{2}$ substituted into (\ref{SSRM}), the maximization of the lower bound of the objective function in (\ref{SSRM}) can be written as (\ref{SSRM_4}) at the top of the next page.
\newcounter{mytempeqncnt}
\begin{figure*}[!t]
\normalsize
\setcounter{equation}{37}
\begin{align}\label{SSRM_4}
&\underset{\mathbf{\mathit{\{\textbf{w}_{i}\},\{\textbf{q}_{l}\}}}}{\mathop{\mathrm{maximize}}} \nonumber\\
&\log\Big(\prod_{i=1}^{M}\frac{\sum\limits_{m=1}^{M}|\textbf{h}^{H}(\theta_{B\!D_i})\textbf{w}_{m}|^{2}+\sum\limits_{l=1}^{L}|\textbf{h}^{H}(\theta_{B\!D_i})\textbf{q}_{l}|^{2}+\sigma_{D}^{2}}
{\sum\limits_{\substack{m=1\\m\neq{i}}}^{M}|\textbf{h}^{H}(\theta_{B\!D_i})\textbf{w}_{m}|^{2}+\sum\limits_{l=1}^{L}|\textbf{h}^{H}(\theta_{B\!D_i})\textbf{q}_{l}|^{2}+\sigma_{D}^{2}}\Big)- \nonumber\\
&\log\!\Big(\prod_{i=1}^{M}\!\underset{k}{\mathop{\max}}\!
\frac{\sum\limits_{m=1}^{M}\!(\textbf{h}^{H}(\hat{\theta}_{B\!E_k})\textbf{w}_m\textbf{w}_m^{H}\textbf{h}(\hat{\theta}_{B\!E_k})\!+\!2\varepsilon_k\|\textbf{w}_m\textbf{w}_m^{H}\textbf{h}(\hat{\theta}_{B\!E_k})\|)
     \!+\!\sum\limits_{l=1}^{L}\!(\textbf{h}^{H}(\hat{\theta}_{B\!E_k})\textbf{q}_l\textbf{q}_l^{H}\textbf{h}(\hat{\theta}_{B\!E_k})\!+\!2\varepsilon_k\|\textbf{q}_l\textbf{q}_l^{H}\textbf{h}(\hat{\theta}_{B\!E_k})\|)\!+\!\sigma_{E}^{2}
     }
{\sum\limits_{\substack{m=1\\m\neq{i}}}^{M}\!(\textbf{h}^{H}(\hat{\theta}_{B\!E_k})\textbf{w}_m\textbf{w}_m^{H}\textbf{h}(\hat{\theta}_{B\!E_k})\!-\!2\varepsilon_k\|\textbf{w}_m\textbf{w}_m^{H}\textbf{h}(\hat{\theta}_{B\!E_k})\|)
     \!+\!\sum\limits_{l=1}^{L}\!(\textbf{h}^{H}(\hat{\theta}_{B\!E_k})\textbf{q}_l\textbf{q}_l^{H}\textbf{h}(\hat{\theta}_{B\!E_k})\!-\!2\varepsilon_k\|\textbf{q}_l\textbf{q}_l^{H}\textbf{h}(\hat{\theta}_{B\!E_k})\|)\!+\!\sigma_{E}^{2}
}\Big).
\end{align}

\setcounter{equation}{38}
\begin{align}\label{SSRM_5}
&\underset{\mathbf{\mathit{\{\textbf{W}_{i}\},\textbf{Q}}}}{\mathop{\mathrm{maximize}}} \nonumber\\
&\log\Big(\prod_{i=1}^{M}\frac{\sum\limits_{m=1}^{M}\mathrm{Tr}\Big(\textbf{H}(\theta_{B\!D_i})\textbf{W}_m\Big)+\mathrm{Tr}\Big(\textbf{H}(\theta_{B\!D_i})\textbf{Q}\Big)+\sigma_{D}^{2}}
{\sum\limits_{\substack{m=1\\m\neq{i}}}^{M}\mathrm{Tr}\Big(\textbf{H}(\theta_{B\!D_i})\textbf{W}_m\Big)+\mathrm{Tr}\Big(\textbf{H}(\theta_{B\!D_i})\textbf{Q}\Big)+\sigma_{D}^{2}}\Big)- \nonumber\\
&\log\!\Big(\prod_{i=1}^{M}\!\underset{k}{\mathop{\max}}\!
\frac{\sum\limits_{m=1}^{M}\!\Big(\mathrm{Tr}(\textbf{H}(\hat{\theta}_{B\!E_k})\textbf{W}_m)\!+\!2\varepsilon_k\|\textbf{W}_m\textbf{h}(\hat{\theta}_{B\!E_k})\|\Big)
     \!+\!\Big(\mathrm{Tr}(\textbf{H}(\hat{\theta}_{B\!E_k})\textbf{Q})\!+\!2\varepsilon_k\|\textbf{Q}\textbf{h}(\hat{\theta}_{B\!E_k})\|\Big)\!+\!\sigma_{E}^{2}
     }
{\sum\limits_{\substack{m=1\\m\neq{i}}}^{M}\!\Big(\mathrm{Tr}(\textbf{H}(\hat{\theta}_{B\!E_k})\textbf{W}_m)\!-\!2\varepsilon_k\|\textbf{W}_m\textbf{h}(\hat{\theta}_{B\!E_k})\|\Big)
     \!+\!\Big(\mathrm{Tr}(\textbf{H}(\hat{\theta}_{B\!E_k})\textbf{Q})\!-\!2\varepsilon_k\|\textbf{Q}\textbf{h}(\hat{\theta}_{B\!E_k})\|\Big)\!+\!\sigma_{E}^{2}
}\Big) \nonumber \\
&\mathrm{s.t.}~\mathrm{Tr}(\textbf{Q})+\sum\limits_{i=1}^{M}\mathrm{Tr}(\textbf{W}_i)\leq{P_t},\nonumber\\
&{\phantom{=}\;\;}~\textbf{W}_i,\textbf{Q}\succeq\textbf{0},\forall{i}.
\end{align}
\hrulefill
\vspace*{4pt}
\end{figure*}
So we have converted (\ref{SSRM}) into a problem to maximize the lower bound of the sum secrecy rate. But (\ref{SSRM_4}) is non-convex and difficult to tackle. To solve this problem, we use the change of variable method in \cite{P.Zhao}. First, the new matrices $\textbf{H}(\theta_{B\!D_i})=\textbf{h}(\theta_{B\!D_i})\textbf{h}^{H}(\theta_{B\!D_i})$,
$\textbf{H}(\hat{\theta}_{B\!E_k})=\textbf{h}(\hat{\theta}_{B\!E_k})\textbf{h}^{H}(\hat{\theta}_{B\!E_k})$ and
the positive semi-definite matrix variables $\textbf{W}_i=\textbf{w}_i\textbf{w}^{H}_i$, $\textbf{Q}=\sum\limits_{l=1}^{L}\textbf{q}_l\textbf{q}^{H}_l$  are introduced to make (\ref{SSRM_4}) more tractable. And (\ref{SSRM_4}) can be rewritten in a simple form as shown in (\ref{SSRM_5}) at the top of this page.

\par
Then, the numerators and denominators of the fractions in the objective function in (\ref{SSRM_5}) are substituted by the exponential variables as follows
\begin{equation}\label{exp_si}
e^{s_i}=\sum\limits_{m=1}^{M}\mathrm{Tr}\Big(\textbf{H}(\theta_{B\!D_i})\textbf{W}_m\Big)+\mathrm{Tr}\Big(\textbf{H}(\theta_{B\!D_i})\textbf{Q}\Big)+\sigma_{D}^{2},
\end{equation}
\begin{equation}\label{exp_ti}
e^{t_i}=\sum\limits_{\substack{m=1\\m\neq{i}}}^{M}\mathrm{Tr}\Big(\textbf{H}(\theta_{B\!D_i})\textbf{W}_m\Big)+\mathrm{Tr}\Big(\textbf{H}(\theta_{B\!D_i})\textbf{Q}\Big)+\sigma_{D}^{2},
\end{equation}
\begin{equation}\label{exp_ak}
\begin{aligned}
e^{a_k}&=\sum\limits_{m=1}^{M}\Big(\mathrm{Tr}(\textbf{H}(\hat{\theta}_{B\!E_k})\textbf{W}_m)+2\varepsilon_k\|\textbf{W}_m\textbf{h}(\hat{\theta}_{B\!E_k})\|\Big)\\
       &\quad+\Big(\mathrm{Tr}(\textbf{H}(\hat{\theta}_{B\!E_k})\textbf{Q})+2\varepsilon_k\|\textbf{Q}\textbf{h}(\hat{\theta}_{B\!E_k})\|\Big)+\sigma_{E}^{2},
\end{aligned}
\end{equation}
\begin{equation}\label{exp_bik}
\begin{aligned}
e^{b_{i,k}}&=\sum\limits_{\substack{m=1\\m\neq{i}}}^{M}\Big(\mathrm{Tr}(\textbf{H}(\hat{\theta}_{B\!E_k})\textbf{W}_m)-2\varepsilon_k\|\textbf{W}_m\textbf{h}(\hat{\theta}_{B\!E_k})\|\Big)\\
             &\quad+\Big(\mathrm{Tr}(\textbf{H}(\hat{\theta}_{B\!E_k})\textbf{Q})-2\varepsilon_k\|\textbf{Q}\textbf{h}(\hat{\theta}_{B\!E_k})\|\Big)+\sigma_{E}^{2}.
\end{aligned}
\end{equation}

\par
Thus we can rewrite the objective function in (\ref{SSRM_5}) as
\begin{equation}
\sum\limits_{i=1}^{M}\big((s_i-t_i)-\underset{k=1,\cdots,K}{\mathop{\max}}(a_k-b_{i,k})\big).
\end{equation}
At the same time, $s_i$, $t_i$, $a_k$, $b_{i,k}$ are constrained by each expression at the right hand sides of (\ref{exp_si}), (\ref{exp_ti}), (\ref{exp_ak}) and (\ref{exp_bik}). Then (\ref{SSRM_5}) can be rewritten as
\begin{subequations}\label{SSRM_6}
\begin{align}
&\underset{\mathbb{S}_1}{\mathop{\mathrm{maximize}}}~ \sum\limits_{i=1}^{M}\big((s_i-t_i)-\underset{k=1,\cdots,K}{\mathop{\max}}(a_k-b_{i,k})\big)\\
&\mathrm{s.t.} \nonumber \\
&\sum\limits_{m=1}^{M}\mathrm{Tr}\Big(\textbf{H}(\theta_{B\!D_i})\textbf{W}_m\Big)+\mathrm{Tr}\Big(\textbf{H}(\theta_{B\!D_i})\textbf{Q}\Big)+\sigma_{D}^{2}\geq{e^{s_i}},\forall{i},\\
&\sum\limits_{\substack{m=1\\m\neq{i}}}^{M}\mathrm{Tr}\Big(\textbf{H}(\theta_{B\!D_i})\textbf{W}_m\Big)+\mathrm{Tr}\Big(\textbf{H}(\theta_{B\!D_i})\textbf{Q}\Big)+\sigma_{D}^{2}\leq{e^{t_i}},\forall{i},\\
&\sum\limits_{m=1}^{M}\Big(\mathrm{Tr}(\textbf{H}(\hat{\theta}_{B\!E_k})\textbf{W}_m)+2\varepsilon_k\|\textbf{W}_m\textbf{h}(\hat{\theta}_{B\!E_k})\|\Big)\nonumber\\
&+\Big(\mathrm{Tr}(\textbf{H}(\hat{\theta}_{B\!E_k})\textbf{Q})+2\varepsilon_k\|\textbf{Q}\textbf{h}(\hat{\theta}_{B\!E_k})\|\Big)+\sigma_{E}^{2}\leq{e^{a_k}},\forall{k},\\
&\sum\limits_{\substack{m=1\\m\neq{i}}}^{M}\Big(\mathrm{Tr}(\textbf{H}(\hat{\theta}_{B\!E_k})\textbf{W}_m)-2\varepsilon_k\|\textbf{W}_m\textbf{h}(\hat{\theta}_{B\!E_k})\|\Big)\nonumber\\
&+\Big(\mathrm{Tr}(\textbf{H}(\hat{\theta}_{B\!E_k})\textbf{Q})-2\varepsilon_k\|\textbf{Q}\textbf{h}(\hat{\theta}_{B\!E_k})\|\Big)+\sigma_{E}^{2}\geq\!{e^{b_{i,k}}},\!\forall{i},\!\forall{k},\\
&\mathrm{Tr}(\textbf{Q})+\sum\limits_{i=1}^{M}\mathrm{Tr}(\textbf{W}_i)\leq{P_t},\\
&\textbf{W}_i,\textbf{Q}\succeq\textbf{0},\forall{i}.
\end{align}
\end{subequations}

\par
Now, the objective function (\ref{SSRM_6}a) is concave, but the constraints (\ref{SSRM_6}c) and (\ref{SSRM_6}d) are non-convex. In order to convert these two constraints into convex ones, we linearize the exponential terms $e^{t_i}$ and $e^{a_k}$ based on the first-order Taylor approximation as
\begin{equation}
e^{t_i}=e^{\bar{t}_i}(t_i-\bar{t}_i+1),
\end{equation}
\begin{equation}
e^{a_k}=e^{\bar{a}_k}(a_k-\bar{a}_k+1),
\end{equation}
where
\begin{equation}\label{bar_t}
\bar{t}_i=\mathrm{ln}\Big(\sum\limits_{\substack{m=1\\m\neq{i}}}^{M}\mathrm{Tr}\Big(\textbf{H}(\theta_{B\!D_i})\textbf{W}_m\Big)+\mathrm{Tr}\Big(\textbf{H}(\theta_{B\!D_i})\textbf{Q}\Big)+\sigma_{D}^{2}\Big),
\end{equation}
\begin{equation}\label{bar_a}
\begin{aligned}
\bar{a}_k&=\mathrm{ln}\Big(\sum\limits_{m=1}^{M}\Big(\mathrm{Tr}(\textbf{H}(\hat{\theta}_{B\!E_k})\textbf{W}_m)+2\varepsilon_k\|\textbf{W}_m\textbf{h}(\hat{\theta}_{B\!E_k})\|\Big)\\
         &\quad+\Big(\mathrm{Tr}(\textbf{H}(\hat{\theta}_{B\!E_k})\textbf{Q})+2\varepsilon_k\|\textbf{Q}\textbf{h}(\hat{\theta}_{B\!E_k})\|\Big)+\sigma_{E}^{2}\Big),
\end{aligned}
\end{equation}
are the  points around which the linearization are made. Thus the problem (\ref{SSRM}) finally becomes
\begin{subequations}\label{SSRM_7}
\begin{align}
&\underset{\mathbb{S}_1}{\mathop{\mathrm{maximize}}}~ \sum\limits_{i=1}^{M}\big((s_i-t_i)-\underset{k=1,\cdots,K}{\mathop{\max}}(a_k-b_{i,k})\big)\\
&\mathrm{s.t.} \nonumber \\
&\sum\limits_{m=1}^{M}\mathrm{Tr}\Big(\textbf{H}(\theta_{B\!D_i})\textbf{W}_m\Big)+\mathrm{Tr}\Big(\textbf{H}(\theta_{B\!D_i})\textbf{Q}\Big)+\sigma_{D}^{2}\geq{e^{s_i}},\forall{i},\\
&\sum\limits_{\substack{m=1\\m\neq{i}}}^{M}\mathrm{Tr}\Big(\textbf{H}(\theta_{B\!D_i})\textbf{W}_m\Big)+\mathrm{Tr}\Big(\textbf{H}(\theta_{B\!D_i})\textbf{Q}\Big)+\sigma_{D}^{2}\nonumber\\
&\leq{e^{\bar{t}_i}(t_i-\bar{t}_i+1)},\forall{i},\\
&\sum\limits_{m=1}^{M}\Big(\mathrm{Tr}(\textbf{H}(\hat{\theta}_{B\!E_k})\textbf{W}_m)+2\varepsilon_k\|\textbf{W}_m\textbf{h}(\hat{\theta}_{B\!E_k})\|\Big)\nonumber\\
&+\Big(\mathrm{Tr}(\textbf{H}(\hat{\theta}_{B\!E_k})\textbf{Q})+2\varepsilon_k\|\textbf{Q}\textbf{h}(\hat{\theta}_{B\!E_k})\|\Big)+\sigma_{E}^{2}\nonumber\\
&\leq{e^{\bar{a}_k}(a_k-\bar{a}_k+1)},\forall{k},\\
&\sum\limits_{\substack{m=1\\m\neq{i}}}^{M}\Big(\mathrm{Tr}(\textbf{H}(\hat{\theta}_{B\!E_k})\textbf{W}_m)-2\varepsilon_k\|\textbf{W}_m\textbf{h}(\hat{\theta}_{B\!E_k})\|\Big)\nonumber\\
&+\Big(\mathrm{Tr}(\textbf{H}(\hat{\theta}_{B\!E_k})\textbf{Q})-2\varepsilon_k\|\textbf{Q}\textbf{h}(\hat{\theta}_{B\!E_k})\|\Big)+\sigma_{E}^{2}\geq\!{e^{b_{i,k}}},\!\forall{i},\!\forall{k},\\
&\mathrm{Tr}(\textbf{Q})+\sum\limits_{i=1}^{M}\mathrm{Tr}(\textbf{W}_i)\leq{P_t},\\
&\textbf{W}_i,\textbf{Q}\succeq\textbf{0},\forall{i}.
\end{align}
\end{subequations}

\par
Now, (\ref{SSRM_7}) is a convex problem and can be solved iteratively by Algorithm 2 using CVX.
\begin{table}[h]
\centering
\begin{tabular}{l}
\hline
\textbf{Algorithm 2} \quad Algorithm for solving the problem (\ref{SSRM_7}) \\
\hline
1:  Given $\{\tilde{\textbf{w}}_i,i=1,\cdots,M\}$ and $\{\tilde{\textbf{q}}_l,l=1,\cdots,L\}$ randomly that\\
\quad are feasible to (\ref{SSRM_7}); \\
2: Set $\tilde{\textbf{W}}_i[0]=\tilde{\textbf{w}}_{i}\tilde{\textbf{w}}_{i}^{H},i=1,\cdots,M$, $\tilde{\textbf{Q}}[0]=\sum\limits_{l=1}^{L}\tilde{\textbf{q}}_{l}\tilde{\textbf{q}}_{l}^{H}$ and set $n$=0; \\
3:  \textbf{Repeat}\\
4:  Substituting $\tilde{\textbf{W}}_i[n]$ and $\tilde{\textbf{Q}}[n]$ into (\ref{bar_t}) and (\ref{bar_a}) yields $\bar{t}_i[n+1]$ \\ \quad and $\bar{a}_k[n+1]$;  \\
5:  Increment $n=n+1$;  \\
6:  Substituting $\bar{t}_i[n]$ and $\bar{a}_k[n]$ into (\ref{SSRM_7}) yields the optimal solution \\ \quad $\tilde{\textbf{W}}_i[n]$ and $\tilde{\textbf{Q}}[n]$;  \\
7:  \textbf{Until} convergence; \\
8:  Obtain $\tilde{\textbf{Q}}$ and $\{\textbf{w}_{i}^{*},i=1,\cdots,M\}$ by decomposition of \\
\quad $\tilde{\textbf{W}}_i[n]=(\textbf{w}_{i}^{*})(\textbf{w}_{i}^{*})^{H}$ for all $i$ in the case of $\mathrm{rank}(\tilde{\textbf{W}}_i[n])=1$; \\
\quad otherwise the randomization technology \cite{M.Zhang} would be utilized to get \\ \quad a rank-one approximation.\\
\hline
\end{tabular}
\end{table}

\par
%The proposed VMD-SSRM method and MAEE-SSRM method achieve better sum secrecy rates at the cost of higher complexity.
Next, we derive the complexity of our two proposed methods.
According to \cite{A.Ben-Tal}, the per-iteration complexity of the proposed VMD-SSRM method and MAEE-SSRM method can be both approximately calculated as
\begin{equation}\label{comp}
\begin{aligned}
&\mathrm{Comp}_{per-iter}=\\
&\mathcal{O}\Big(\sqrt{(KM+K+2M+1)(2N+2)+KM+(M+1)N}\\
&n_o[n_o^2+n_o((KM+K+2M+1)(2N+2)^2+KM^2+\\
&(M+1)N^2)+(KM+K+2M+1)(2N+2)^3+KM^3+\\
&(M+1)N^3]\mathrm{ln}(\frac{1}{\epsilon})\Big),
\end{aligned}
\end{equation}
where $N$, $M$ and $K$ are the number of transmit antennas at the BS, the number of destination users and the number of eavesdroppers, respectively. $n_o=(K+3)M+K+(M+1)N^2$ is the dimension of real design variables, and $\epsilon$ is a computational accuracy.
Then the complexity of VMD-SSRM method is $\mathrm{Comp}_{per-iter}\cdot I_{V\!M\!D_{max}}$ while the complexity of MAEE-SSRM method is $\mathrm{Comp}_{per-iter}\cdot I_{M\!A\!E\!E_{max}}$, where $I_{V\!M\!D_{max}}$ and $I_{M\!A\!E\!E_{max}}$ are the iteration times of VMD-SSRM method and MAEE-SSRM method, respectively.

\par
According to (\ref{comp}), the per-iteration complexity of VMD-SSRM method and MAEE-SSRM method increases with the increase of $N$, $M$ and $K$.
When $N$, $M$ and $K$ are small, for example, $N=4$, $M=2$, $K=1$, the complexity of proposed VMD-SSRM method and MAEE-SSRM method is slightly higher than that of ZF method and SLNR method.

\section{SIMULATION AND ANALYSIS}
\par
In this section, we present the simulation results to evaluate the performance of our proposed VMD-SSRM method and MAEE-SSRM method.
Our methods are compared to ZF method and SLNR method.
The two methods are based on the schemes in \cite{M.Alageli} and \cite{F.Shu}, respectively, but the eavesdropping channels in these two methods are estimated in the same way as our VMD-SSRM method.
The simulation parameters are set as shown in Table I, unless otherwise stated.
\begin{table}[!h]
\renewcommand{\arraystretch}{1.3}
\caption{Simulation Parameters Setting}
\label{simulation parameters}
\centering
\begin{tabular}{|c|c|}
\hline
Parameters & Values  \\ \hline
\makecell{The number of antennas at the BS ($N$)} & \makecell{$6$} \\ \hline
\makecell{The number of destination users ($M$)} & \makecell{$2$} \\ \hline
\makecell{The number of eavesdroppers ($K$)} & \makecell{$4$} \\ \hline
\makecell{The maximum angle estimation error toward each \\ eavesdropper ($\Delta\theta_{\mathrm{max}}$)} & \makecell{$6^{\circ}$} \\ \hline
\makecell{The mean of the angle estimation error ($\mu$)} & \makecell{$0$} \\ \hline
\makecell{The concentration parameter of the angle \\ estimation error ($\kappa$)} & \makecell{$100$} \\ \hline
\makecell{The power of receiver noise ($\sigma_D^2$, $\sigma_E^2$)} & \makecell{$-30\mathrm{dBm}$} \\ \hline
\makecell{The transmit power of the BS ($P_t$)} & \makecell{$40\mathrm{dBm}$} \\ \hline
\makecell{The direction angles toward the destination users \\ ($\{\theta_{B\!D_1},\theta_{B\!D_2}\}$)} & \makecell{$\{\frac{\pi}{6},\frac{\pi}{3}\}$} \\ \hline
\makecell{The direction angles toward the eavesdroppers \\ ($\{\theta_{B\!E_1},\theta_{B\!E_2},\theta_{B\!E_3},\theta_{B\!E_4}\}$)} & \makecell{$\{-\frac{\pi}{12},\frac{\pi}{12},\frac{\pi}{4},\frac{5\pi}{12}\}$} \\ \hline
\end{tabular}
\end{table}

\begin{figure}[!ht]
\centering
\includegraphics[width=0.45\textwidth]{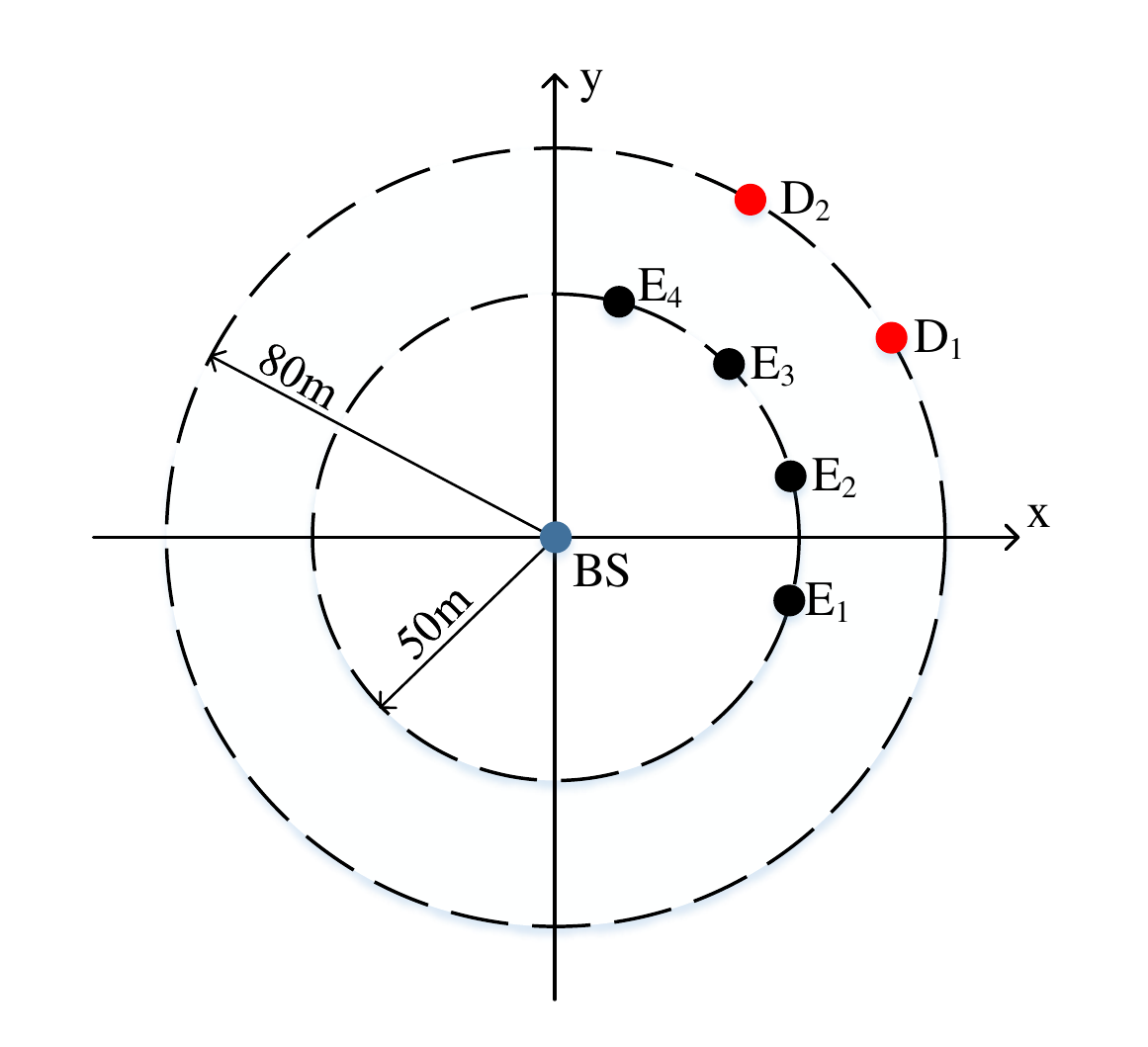}   %2.5in
\caption{The locations of the BS, destination users and eavesdroppers.}
%\caption{Sum secrecy rate versus the transmit power of the BS. The secrecy rate comparison of our proposed SINR maximization, zero-forcing in \cite{R11} and QoSD in \cite{R20}, where $M=8$, $N=4$, $P_i=2.5mW$ and $\sigma^{2}=0.01$ }
\label{fig3}
\end{figure}

The distance between the BS and each destination user is set to 80m, while the distance between the BS and each eavesdropper is set to 50m.
Fig. \ref{fig3} shows the locations of the BS, destination users and eavesdroppers in the Cartesian coordinate system.

\begin{figure}[!ht]
\centering
\includegraphics[width=0.45\textwidth]{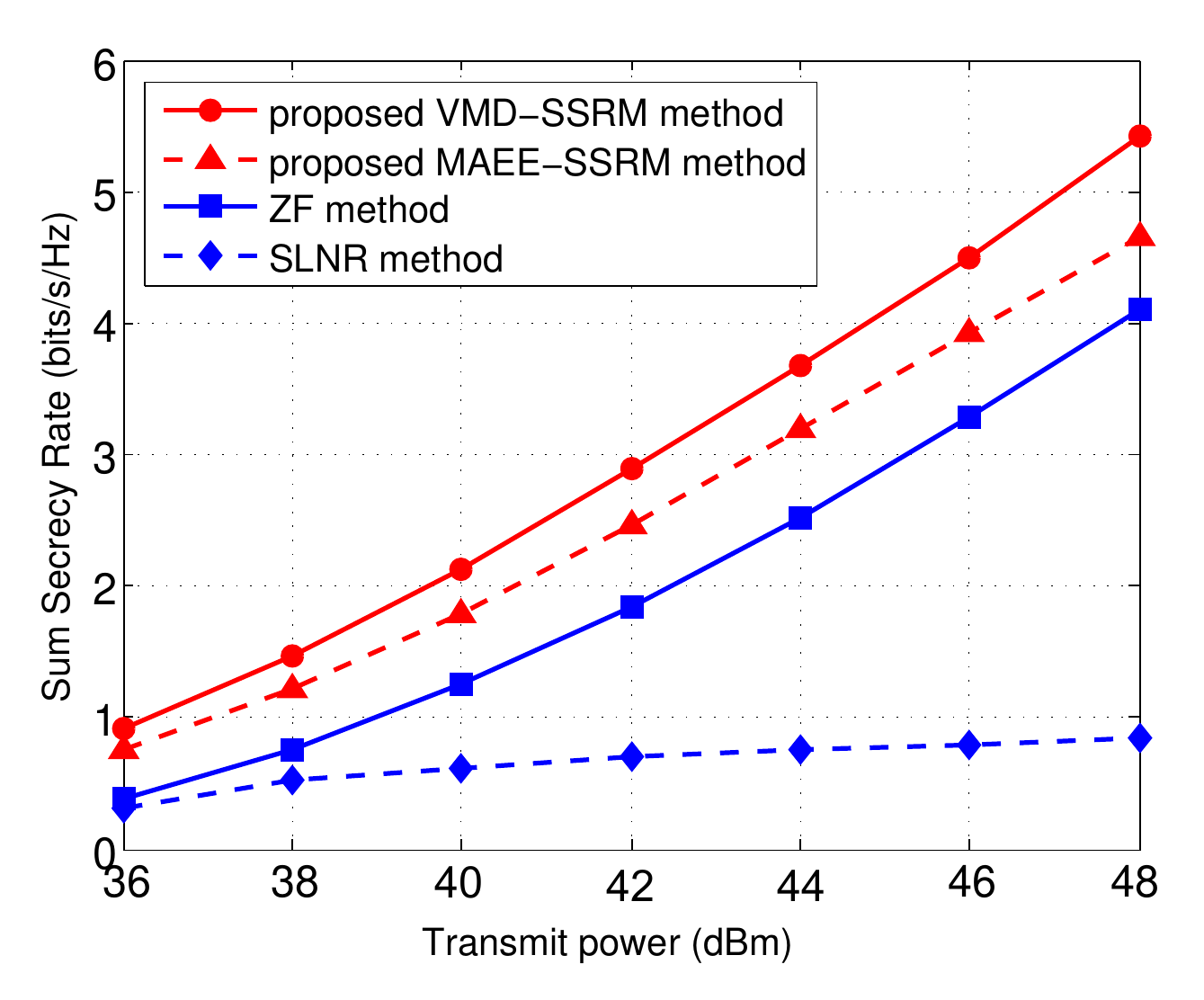}   %2.5in
\caption{Sum secrecy rate versus the transmit power of the BS for $N=6$, $M=2$, $K=4$, $\Delta\theta_{\mathrm{max}}=6^{\circ}$.}
\label{fig4}
\end{figure}

\par
Fig. \ref{fig4} shows how the sum secrecy rate changes with the transmit power of the BS.
It can be observed that the sum secrecy rate grows with the increase of the transmit power at the BS for all four methods.
The reason is explained as follows.
When the transmit power of the BS increases, the destination user's SINR increases more than the eavesdropper's SINR, because these four methods all try to provide the eavesdroppers less confidential signal and more AN by designing the best signal beamforming matrix and AN beamforming matrix, thus limiting the eavesdropper's SINR.
It can also be observed that our VMD-SSRM method achieves better sum secrecy rate than our MAEE-SSRM method.
The reason is that MAEE-SSRM method aims to maximize the system performance in the worst case when the estimation error of direction angle toward each eavesdropper is the largest, while VMD-SSRM method can deal with more cases because most angle estimation errors are taken into consideration through expectation computation.
Moreover, the sum secrecy rate of SLNR method increases slowly.
The reason is explained as follows.
As in \cite{F.Shu}, SLNR method fixes the power allocation factors for both confidential signal and AN, and the power allocation factor for confidential signal is much higher than that of AN.
So when transmit power of the BS increases, the power allocated to confidential signal is much more than that of AN, providing the eavesdroppers more chance to decipher confidential signals.
So the sum secrecy rate of SLNR method increases very slowly.

\begin{figure}[!ht]
\centering
\includegraphics[width=0.45\textwidth]{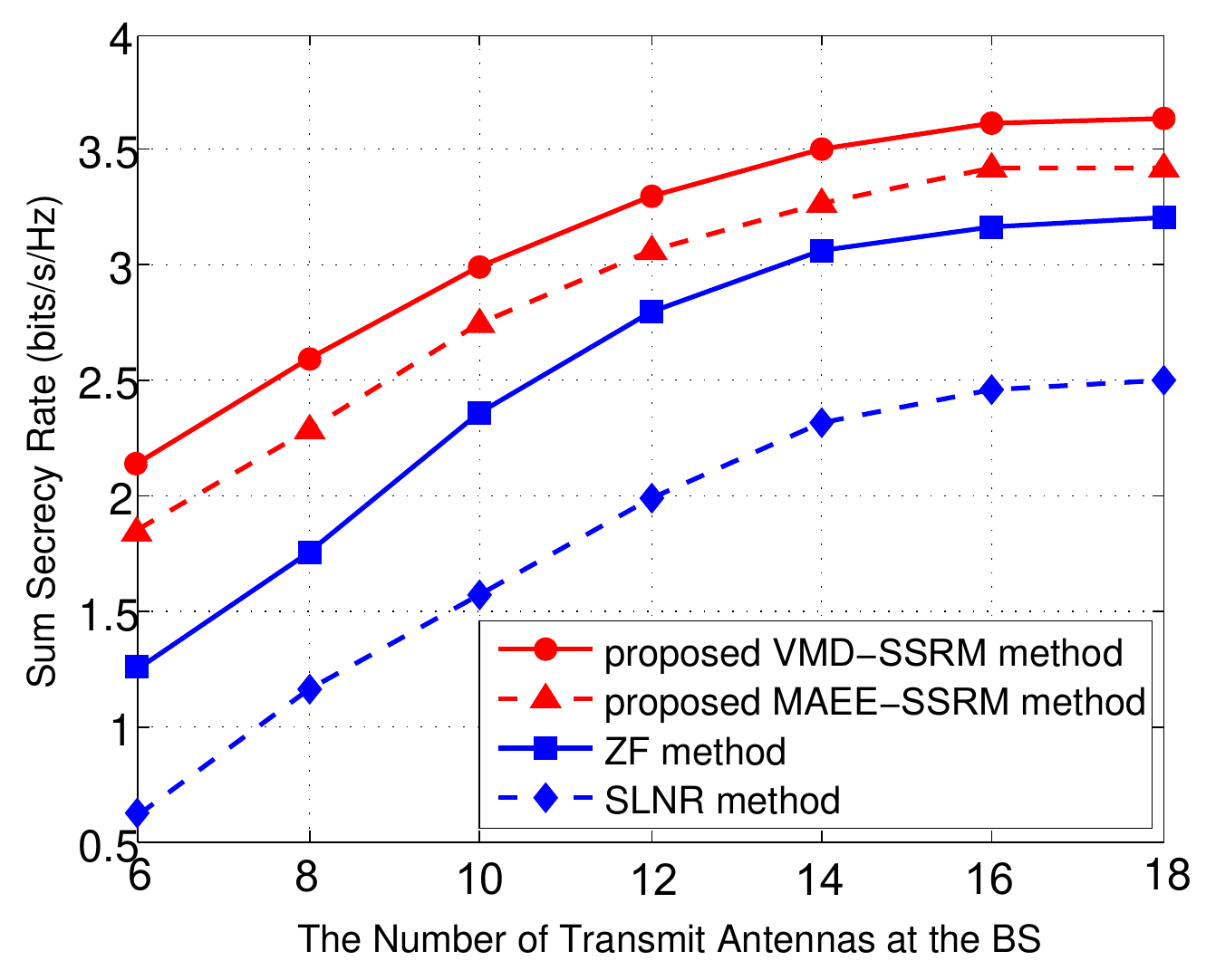}   %2.5in
\caption{Sum secrecy rate versus the number of transmit antennas at the BS for $M=2$, $K=4$, $P_t=40\mathrm{dBm}$, $\Delta\theta_{\mathrm{max}}=6^{\circ}$.}
\label{fig5}
\end{figure}

\par
Fig. \ref{fig5} compares the sum secrecy rate against different number of transmit antennas at the BS.
It can be seen that when the number of transmit antennas increases, the sum secrecy rates of four methods also increase.
The reason is explained as follows.
These four methods all try to obtain the optimal sum secrecy rate by designing the optimal signal beamforming vectors and AN beamforming vectors.
On the other hand, when the BS has more antennas, the beams generated by the BS can be narrower.
So the signal beams will be more precisely directed toward destination users and the AN beams will also be more precisely directed toward eavesdroppers.
It can also be observed that when the number of transmit antennas exceeds some value (such as 16), the sum secrecy rates of four methods increase slowly.
The reason can be explained as follows.
When the number of transmit antennas exceeds 16, the dimension of the beamforming vectors is already high enough to concentrate the confidential signal on the corresponding desired user as well as concentrate the AN on eavesdroppers.
Since the system secrecy performance is already optimized, merely increasing the number of transmit antennas cannot effectively improve the system secrecy performance any more.

\begin{figure}[!ht]
\centering
\includegraphics[width=0.45\textwidth]{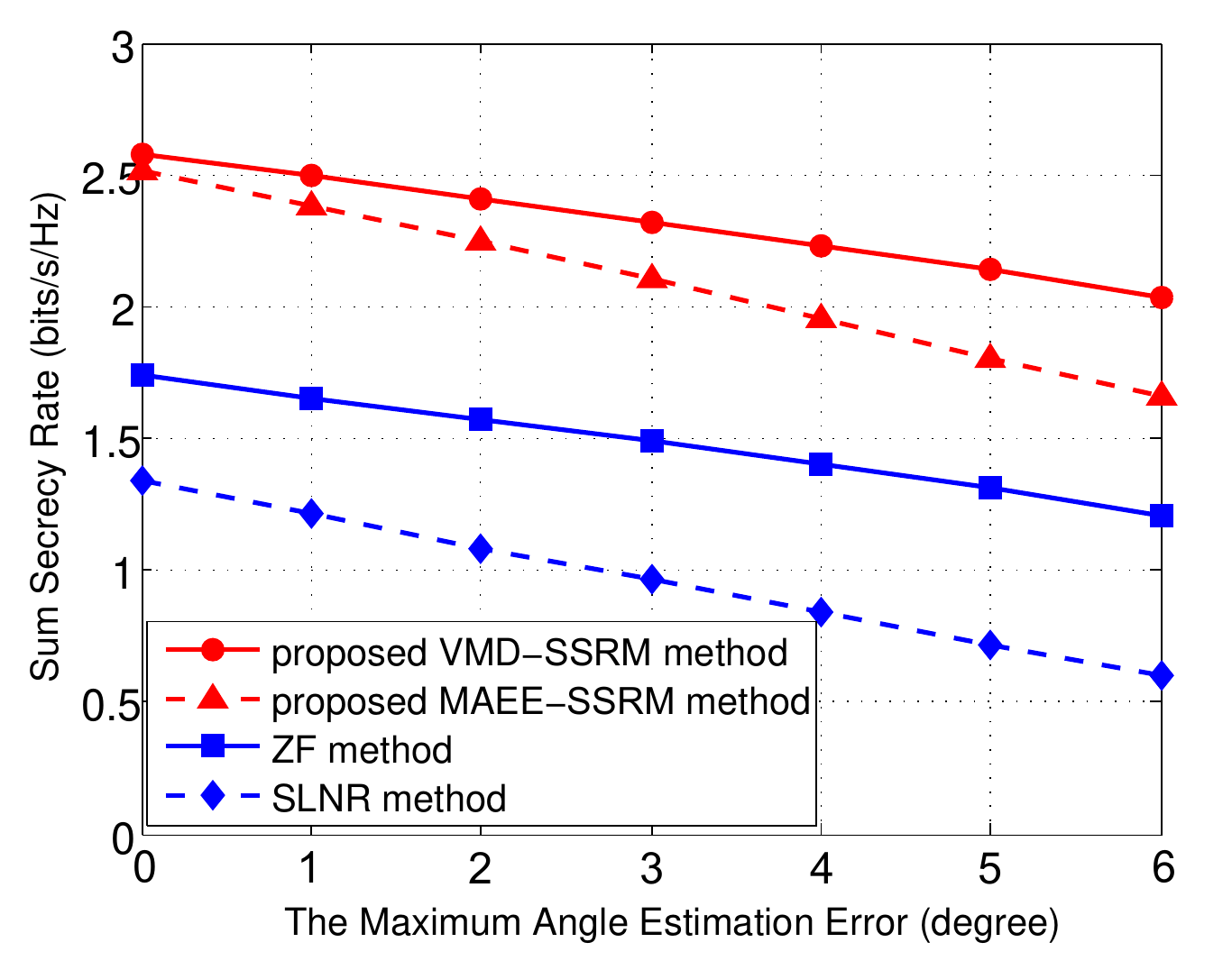}   %2.5in
\caption{Sum secrecy rate versus the maximum angle estimation error for $N=6$, $M=2$, $K=4$, $P_t=40\mathrm{dBm}$.}
\label{fig6}
\end{figure}

\par
In Fig. \ref{fig6}, fixing the transmit power of the BS, we investigate the impact of the maximum angle estimation error $\Delta\theta_{\mathrm{max}}$ on the sum secrecy rate.
With the increase of $\Delta\theta_{\mathrm{max}}$, the sum secrecy rates of the four methods degrade slowly.
The reason is that these four methods all take the estimation error of direction angle toward each eavesdropper into consideration when constructing the optimization problems.
Similar to our VMD-SSRM method, the ZF method and the SLNR method both obtain the expectation of eavesdropping channel related coefficient on the condition of Von-Mises-distributed direction angle estimation error.
As such, they are robust against the effects of the direction angle estimation errors.
Moreover, our proposed VMD-SSRM method and MAEE-SSRM method obviously outperform ZF method and SLNR method.
The main reason is that both VMD-SSRM method and MAEE-SSRM method are aiming to directly maximize the sum secrecy rate, while ZF method and SLNR method do not take the sum secrecy rate as the direct optimization objective.
The optimization objective of ZF method is to maximize the confidential signal for each destination user whist nullify signal interference at each destination user.
On the other hand, SLNR method aims to minimize the confidential signal leakage to other users including all eavesdroppers and meanwhile to maximize AN power leakage to all eavesdroppers \cite{F.Shu}.

\begin{figure}[!ht]
\centering
\includegraphics[width=0.45\textwidth]{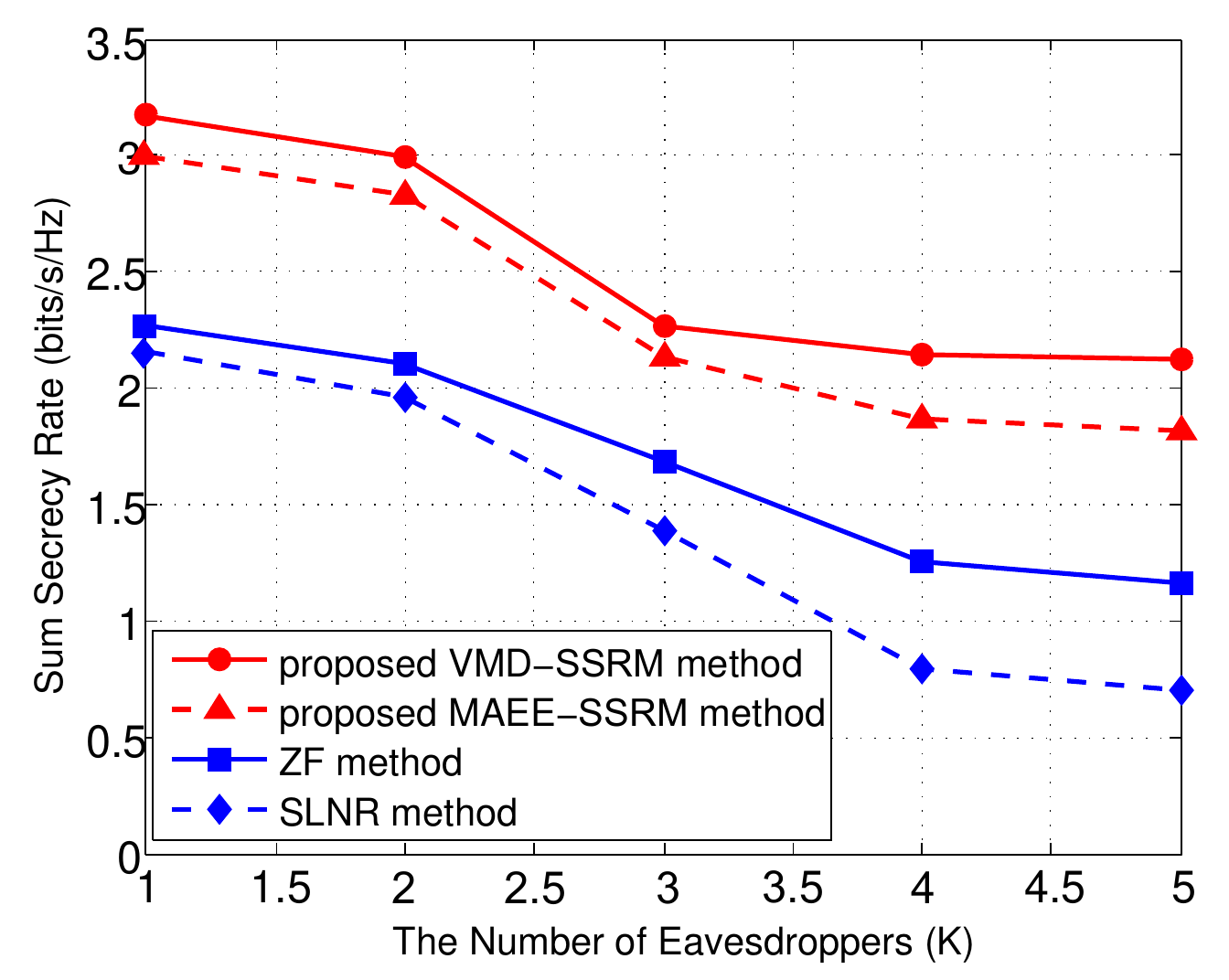}   %2.5in
\caption{Sum secrecy rate versus the number of the eavesdroppers for $N=6$, $M=2$, $\{\theta_{B\!E_1},\theta_{B\!E_2},\theta_{B\!E_3},\theta_{B\!E_4}, \theta_{B\!E_5}\}=\{-\frac{\pi}{12},\frac{\pi}{12},\frac{\pi}{4},\frac{5\pi}{12}, \frac{7\pi}{12}\}$, $P_t=40\mathrm{dBm}$, $\Delta\theta_{\mathrm{max}}=6^{\circ}$.}
\label{fig7}
\end{figure}

\par
Fig. \ref{fig7} shows how the sum secrecy rate changes with the number of the eavesdroppers.
It can be observed that the sum secrecy rates of four methods decrease when the number of the eavesdroppers increases.
The reason is explained as follows.
When there are more eavesdroppers, the confidential signal will be more likely to be wiretapped by the eavesdroppers.
Although the AN is used to interfere the eavesdroppers, since the BS has only 6 transmit antennas, the BS cannot generate enough AN beams to interfere the additional eavesdroppers.
It can also be observed that when the number of the eavesdroppers exceeds 4, the sum secrecy rates of four methods decrease slowly.
The reason is that the fifth eavesdropper $E_5$ (as well as subsequent eavesdroppers) are set to be far away from the destination users.
Thus even if the number of the eavesdroppers increases, the additional eavesdroppers can hardly capture confidential signals, resulting in small effect on the sum secrecy rate.

\section{Conclusion}
In this paper, we have investigated the robust beamforming matrix design for sum secrecy
rate maximization based on directional modulation in multi-user systems.
It was assumed that the BS has the imperfect knowledge of the direction angle toward each eavesdropper and the direction angle estimation errors follow the Von Mises distribution.
Following this assumption, we first proposed a VMD-SSRM method to maximize the sum secrecy rate.
In addition, to optimize the sum secrecy rate in the case of the worst estimation error of direction angle toward each eavesdropper, we proposed a MAEE-SSRM method.
From the analysis and the simulations, our VMD-SSRM method and MAEE-SSRM method achieve better sum secrecy rate than ZF method and SLNR method.
Moreover, the sum secrecy rate performance of the proposed VMD-SSRM method is better than that of the proposed MAEE-SSRM method.

\appendices
\section*{Appendix A}
\renewcommand\theequation{A1}
$H_{B\!E_k}(u,v)$ can be expressed as
\begin{equation}\label{H_BEk_pq_11}
H_{B\!E_k}(u,v)=\mathbb{E}\Big\{\textbf{h}_u(\hat{\theta}_{B\!E_k}+\Delta\theta_{B\!E_k})\textbf{h}_v^{H}(\hat{\theta}_{B\!E_k}+\Delta\theta_{B\!E_k})\Big\}.
\end{equation}
According to (\ref{h_theta_mn}), (\ref{H_BEk_pq_11}) can be written as
\renewcommand\theequation{A2}
\begin{equation}\label{H_BEk_pq_12}
\begin{aligned}
H_{B\!E_k}(u,v)&=\frac{g_{B\!E_k}}{N}\int_{-\Delta\theta_{\mathrm{max}}}^{\Delta\theta_{\mathrm{max}}}e^{j\frac{2\pi(v-u)l}{\lambda}\cos(\hat{\theta}_{B\!E_k}+\Delta\theta_{B\!E_k})}\\
&\quad f(\Delta\theta_{B\!E_k}\!)d(\Delta\theta_{B\!E_k}).
\end{aligned}
\end{equation}
The exponential function in (\ref{H_BEk_pq_12}) can be expanded by using Euler's formula as
\renewcommand\theequation{A3}
\begin{equation}\label{H_BEk_pq_13}
\begin{aligned}
&H_{B\!E_k}(u,v)=\\
&\frac{g_{B\!E_k}}{N}\int_{-\Delta\theta_{\mathrm{max}}}^{\Delta\theta_{\mathrm{max}}}\mathrm{exp}\Big\{j\alpha_{uv}\Big(\cos(\hat{\theta}_{B\!E_k})\cos(\Delta\theta_{B\!E_k})\\
&-\sin(\hat{\theta}_{B\!E_k})\sin(\Delta\theta_{B\!E_k})\Big)\Big\}f(\Delta\theta_{B\!E_k}\!)d(\Delta\theta_{B\!E_k}),
\end{aligned}
\end{equation}
where $\alpha_{uv}=\frac{2\pi(v-u)l}{\lambda}$.
It is assumed that $\Delta\theta_{\mathrm{max}}$  is so small that it approaches zero. The trigonometric functions of the direction angle estimation error in (\ref{H_BEk_pq_13}) can be expanded by using second-order Taylor expansion.
Thus we can get the following approximate expressions
\renewcommand\theequation{A4}
\begin{equation}\label{sin_appro}
\sin(\Delta\theta_{B\!E_k})\approx\Delta\theta_{B\!E_k},
\end{equation}
\renewcommand\theequation{A5}
\begin{equation}\label{cos_appro}
\cos(\Delta\theta_{B\!E_k})\approx1-\frac{\Delta\theta_{B\!E_k}^{2}}{2}.
\end{equation}
\noindent Substituting (\ref{sin_appro}) and (\ref{cos_appro}) into (\ref{H_BEk_pq_13}), we have the expression of $H_{B\!E_k}(u,v)$ as
\renewcommand\theequation{A6}
\begin{equation}\label{H_BEk_pq_21}
\begin{aligned}
&H_{B\!E_k}(u,v)=\\
&\frac{g_{B\!E_k}}{N}e^{j\alpha_{u\!v}\cos(\hat{\theta}_{B\!E_k})}\int_{-\Delta\theta_{\mathrm{max}}}^{\Delta\theta_{\mathrm{max}}}\!\mathrm{exp}\!\Big\{\!-j\alpha_{uv}\Big(\!\cos(\hat{\theta}_{B\!E_k})\!\frac{\Delta\theta_{B\!E_k}^{2}}{2}\\ &+\sin(\hat{\theta}_{B\!E_k})\Delta\theta_{B\!E_k}\Big)\Big\}f(\Delta\theta_{B\!E_k}\!)d(\Delta\theta_{B\!E_k}).
\end{aligned}
\end{equation}
According to Euler's formula, (\ref{H_BEk_pq_21}) can be further written as
\renewcommand\theequation{A7}
\begin{equation}\label{H_BEk_pq_22}
\begin{aligned}
H_{B\!E_k}(u,v)=&\hat{H}_{B\!E_k}\!(u,v)\!\int_{-\Delta\theta_{\mathrm{max}}}^{\Delta\theta_{\mathrm{max}}}\!\Big(\cos(\xi_{u,v}^{k})\!-\!j\sin(\xi_{u,v}^{k})\!\Big)\\
&f(\Delta\theta_{B\!E_k}\!)d(\Delta\theta_{B\!E_k}),
\end{aligned}
\end{equation}
where
\renewcommand\theequation{A8}
\begin{equation}
\hat{H}_{B\!E_k}\!(u,v)=\frac{g_{B\!E_k}}{N}\!e^{j\!\alpha_{uv}\!\cos(\hat{\theta}_{B\!E_k})},
\end{equation}
denotes the $u$-th row and the $v$-th column entry of $\textbf{h}(\hat{\theta}_{B\!E_k})\textbf{h}^{H}(\hat{\theta}_{B\!E_k})$, and
\renewcommand\theequation{A9}
\begin{equation}\label{xi}
\xi_{u,v}^{k}=\alpha_{uv}\Big(\!\cos(\hat{\theta}_{B\!E_k})\!\frac{\Delta\theta_{B\!E_k}^{2}}{2}+\sin(\hat{\theta}_{B\!E_k})\Delta\theta_{B\!E_k}\Big).
\end{equation}
For simplify, (\ref{H_BEk_pq_22}) can be written as
\renewcommand\theequation{A10}
\begin{equation}\label{H_BEk_pq_23}
H_{B\!E_k}(u,v)=\Upsilon_{1k}(u,v)-j\Upsilon_{2k}(u,v),
\end{equation}
where
\renewcommand\theequation{A11}
\begin{equation}\label{Upsilon_1}
\Upsilon_{1k}(u,v)\!=\!\hat{H}_{B\!E_k}\!(u,v)\!\int_{-\Delta\theta_{\mathrm{max}}}^{\Delta\theta_{\mathrm{max}}}\!\cos(\xi_{u,v}^{k})\!f(\Delta\theta_{B\!E_k}\!)d(\Delta\theta_{B\!E_k}),
\end{equation}
\renewcommand\theequation{A12}
\begin{equation}\label{Upsilon_2}
\Upsilon_{2k}(u,v)\!=\!\hat{H}_{B\!E_k}\!(u,v)\!\int_{-\Delta\theta_{\mathrm{max}}}^{\Delta\theta_{\mathrm{max}}}\!\sin(\xi_{u,v}^{k})\!f(\Delta\theta_{B\!E_k}\!)d(\Delta\theta_{B\!E_k}).
\end{equation}

\par
Now the task is to derive the analytic expressions of $\Upsilon_{1k}(u,v)$ and $\Upsilon_{2k}(u,v)$.
For this purpose, we first expand the Cosine function in (\ref{Upsilon_1}) into the following form
\renewcommand\theequation{A13}
\begin{equation}\label{cos}
\begin{aligned}
&\cos(\xi_{u,v}^{k})=\\
&\cos\Big(\alpha_{uv}\cos(\hat{\theta}_{B\!E_k})\frac{\Delta\theta_{B\!E_k}^{2}}{2}\Big)\cos\Big(\alpha_{uv}\sin(\hat{\theta}_{B\!E_k})\Delta\theta_{B\!E_k}\Big)\\
&-\sin\Big(\alpha_{uv}\cos(\hat{\theta}_{B\!E_k})\frac{\Delta\theta_{B\!E_k}^{2}}{2}\Big)\sin\Big(\alpha_{uv}\sin(\hat{\theta}_{B\!E_k})\Delta\theta_{B\!E_k}\Big).
\end{aligned}
\end{equation}
Then the second-order Taylor series is used to approximate each term in (\ref{cos}), and (\ref{cos}) can be rewritten as
\renewcommand\theequation{A14}
\begin{equation}\label{cos_1}
\begin{aligned}
&\cos(\xi_{u,v}^{k})=\\
&\Big(1-\frac{\alpha_{uv}^{2}\cos^{2}(\hat{\theta}_{B\!E_k})\Delta\theta_{B\!E_k}^{4}}{8}\Big)
\Big(1-\frac{\alpha_{uv}^{2}\sin^{2}(\hat{\theta}_{B\!E_k})\Delta\theta_{B\!E_k}^{2}}{2}\Big)\\
&-\Big(\alpha_{uv}\cos(\hat{\theta}_{B\!E_k})\frac{\Delta\theta_{B\!E_k}^{2}}{2}\Big)
\Big(\alpha_{uv}\sin(\hat{\theta}_{B\!E_k})\Delta\theta_{B\!E_k}\Big).
\end{aligned}
\end{equation}
Expand (\ref{cos_1}) as a sum of several terms and omit the higher order terms about $\Delta\theta_{B\!E_k}$, then (\ref{cos_1}) can be further written as
\renewcommand\theequation{A15}
\begin{equation}\label{cos_11}
\begin{aligned}
\cos(\xi_{u,v}^{k})&\approx1-\frac{\alpha_{uv}^{2}\sin^{2}(\hat{\theta}_{B\!E_k})\Delta\theta_{B\!E_k}^{2}}{2}\\
&\quad -\alpha_{uv}^{2}\sin(\hat{\theta}_{B\!E_k})\cos(\hat{\theta}_{B\!E_k})\frac{\Delta\theta_{B\!E_k}^{3}}{2}.
\end{aligned}
\end{equation}

\par
Since the direction angle estimation error $\Delta\theta_{B\!E_k}$ is assumed to be very small, $\mu$ is also a small value near zero. Therefore, we can expand the trigonometric function in the Von Mises PDF as
\renewcommand\theequation{A16}
\begin{equation}\label{cos_2}
\cos(\Delta\theta_{B\!E_k}-\mu)\approx1-\frac{1}{2}(\Delta\theta_{B\!E_k}-\mu)^{2}.
\end{equation}
Combining (\ref{Upsilon_1}), (\ref{cos_11}) and (\ref{cos_2}), we can obtain the expression of $\Upsilon_{1k}(u,v)$ at the top of the next page.
\newcounter{mytempeqncnt1}
\begin{figure*}[!t]
\normalsize
\renewcommand\theequation{A17}
\begin{align}\label{Gamma_1k}
\Upsilon_{1k}(u,v)&\approx\hat{H}_{B\!E_k}(u,v)\int_{-\Delta\theta_{\mathrm{max}}}^{\Delta\theta_{\mathrm{max}}}\Big(1-\frac{\alpha_{uv}^{2}\sin^{2}(\hat{\theta}_{B\!E_k})}{2}\Delta\theta_{B\!E_k}^{2}-\frac{\alpha_{uv}^{2}\sin(\hat{\theta}_{B\!E_k})\cos(\hat{\theta}_{B\!E_k})}{2}\Delta\theta_{B\!E_k}^{3}\Big)f(\Delta\theta_{B\!E_k})d(\Delta\theta_{B\!E_k}) \nonumber \\
&\overset{(b)}{\approx}\frac{\hat{H}_{B\!E_k}(u,v)e^\kappa}{2\pi I_0(\kappa)}\Big\{\sqrt{\frac{\pi}{2\kappa}}\Big(\mathrm{erf}(\sqrt{\frac{\kappa}{2}}\Delta\theta_1)+\mathrm{erf}(\sqrt{\frac{\kappa}{2}}\Delta\theta_2)\Big) \nonumber\\
&\quad-\frac{\alpha_{uv}^{2}\sin^{2}(\hat{\theta}_{B\!E_k})}{2\kappa}\Big(\frac{(1+\kappa\mu^{2})\sqrt{\pi}}{\sqrt{2\kappa}}\Big(\mathrm{erf}(\sqrt{\frac{\kappa}{2}}\Delta\theta_1)+\mathrm{erf}(\sqrt{\frac{\kappa}{2}}\Delta\theta_2)\Big)-\Delta\theta_{2}e^{-\frac{\kappa}{2}\Delta\theta_{1}^{2}}-\Delta\theta_{1}e^{-\frac{\kappa}{2}\Delta\theta_{2}^{2}}\Big)\nonumber \\
&\quad-\frac{\alpha_{uv}^{2}\sin(\hat{\theta}_{B\!E_k})\cos(\hat{\theta}_{B\!E_k})}{2\kappa}\Big(\frac{(3+\kappa\mu^{2})\mu\sqrt{\pi}}{\sqrt{2\kappa}}\Big(\mathrm{erf}(\sqrt{\frac{\kappa}{2}}\Delta\theta_1)+\mathrm{erf}(\sqrt{\frac{\kappa}{2}}\Delta\theta_2)\Big)\nonumber\\
&\quad+(\frac{2}{\kappa}+\Delta\theta_{1}^{2}-3\mu)e^{-\frac{\kappa}{2}\Delta\theta_{1}^{2}}-(\frac{2}{\kappa}+\Delta\theta_{2}^{2}+3\mu)e^{-\frac{\kappa}{2}\Delta\theta_{2}^{2}}\Big)\Big\}.
\end{align}

\renewcommand\theequation{A24}
\begin{align}\label{Gamma_2k}
\Upsilon_{2k}(u,v)&\approx\hat{H}_{B\!E_k}(u,v)\int_{-\Delta\theta_{\mathrm{max}}}^{\Delta\theta_{\mathrm{max}}}\Big(\frac{\alpha_{uv}\cos(\hat{\theta}_{B\!E_k})}{2}\Delta\theta_{B\!E_K}^2+\alpha_{uv}\sin(\hat{\theta}_{B\!E_k})\Delta\theta_{B\!E_K}\Big)f(\Delta\theta_{B\!E_k})d(\Delta\theta_{B\!E_k}) \nonumber \\
&\approx\frac{\hat{H}_{B\!E_k}(u,v)e^\kappa}{2\pi I_0(\kappa)}\Big\{\frac{\alpha_{uv}\cos(\hat{\theta}_{B\!E_k})}{2\kappa}\Big(\frac{(1+\kappa\mu^2)\sqrt{\pi}}{\sqrt{2\kappa}}\Big(\mathrm{erf}(\sqrt{\frac{\kappa}{2}}\Delta\theta_1)+\mathrm{erf}(\sqrt{\frac{\kappa}{2}}\Delta\theta_2)\Big)-\Delta\theta_{2}e^{-\frac{\kappa}{2}\Delta\theta_{1}^{2}}-\Delta\theta_{1}e^{-\frac{\kappa}{2}\Delta\theta_{2}^{2}}\Big)\nonumber\\
&\quad+\alpha_{uv}\sin(\hat{\theta}_{B\!E_k})\Big(\mu\sqrt{\frac{\pi}{2\kappa}}\Big(\mathrm{erf}(\sqrt{\frac{\kappa}{2}}\Delta\theta_1)+\mathrm{erf}(\sqrt{\frac{\kappa}{2}}\Delta\theta_2)\Big)+\frac{1}{\kappa}(e^{-\frac{\kappa}{2}\Delta\theta_{1}^{2}}-e^{-\frac{\kappa}{2}\Delta\theta_{2}^{2}})\Big)\Big\}.
\end{align}
\hrulefill
\vspace*{4pt}
\end{figure*}

\par
Note that in (\ref{Gamma_1k}), $\Delta\theta_{1}=\Delta\theta_{\mathrm{max}}+\mu$, $\Delta\theta_{2}=\Delta\theta_{\mathrm{max}}-\mu$ and the step (b) results from the following equations \cite{I.S.Gradshteyn}
\renewcommand\theequation{A18}
\begin{equation}
\int_{0}^{x}e^{-q^{2}t^{2}}dt=\frac{\sqrt{\pi}}{2q}\mathrm{erf}(qx),
\end{equation}
\renewcommand\theequation{A19}
\begin{equation}
\int_{0}^{x}te^{-q^{2}t^{2}}dt=\frac{1}{2q^{2}}(1-e^{-q^{2}x^{2}}),
\end{equation}
\renewcommand\theequation{A20}
\begin{equation}
\int_{0}^{x}t^{2}e^{-q^{2}t^{2}}dt=\frac{1}{2q^{3}}\Big(\frac{\sqrt{\pi}}{2}\mathrm{erf}(qx)-qxe^{-q^{2}x^{2}}\Big),
\end{equation}
\renewcommand\theequation{A21}
\begin{equation}
\int_{0}^{x}t^{3}e^{-q^{2}t^{2}}dt=\frac{1}{2q^{4}}\Big(1-(1+q^{2}x^{2})e^{-q^{2}x^{2}}\Big),
\end{equation}
where $\mathrm{erf}(x)$ represents the error function defined as
\renewcommand\theequation{A22}
\begin{equation}
\mathrm{erf}(x)=\frac{2}{\sqrt{\pi}}\int_{0}^{x}e^{-t^{2}}dt.
\end{equation}

\par
Similarly, $\sin(\xi_{u,v}^{k})$ in (\ref{Upsilon_2}) can be approximately expressed as
\renewcommand\theequation{A23}
\begin{equation}\label{sin}
\begin{aligned}
&\sin(\xi_{u,v}^{k})\approx\\
%&\sin\Big(\alpha_{uv}\cos(\hat{\theta}_{B\!E_k})\frac{\Delta\theta_{B\!E_k}^{2}}{2}\Big)\cos\Big( \alpha_{uv}\sin(\hat{\theta}_{B\!E_k})\Delta\theta_{B\!E_k}\Big) \\
%&+\cos\Big(\alpha_{uv}\cos(\hat{\theta}_{B\!E_k})\frac{\Delta\theta_{B\!E_k}^{2}}{2}\Big)\sin\Big(\alpha_{uv}\sin(\hat{\theta}_{B\!E_k})\Delta\theta_{B\!E_k}\Big) \\
&\Big(\frac{\alpha_{uv}\cos(\hat{\theta}_{B\!E_k})}{2}\Delta\theta_{B\!E_k}^{2}\Big)\Big(1-\frac{\alpha_{uv}^{2}\sin^{2}(\hat{\theta}_{B\!E_k})\Delta\theta_{B\!E_k}^{2}}{2}\Big)\\
&+\Big(1-\frac{\alpha_{uv}^{2}\cos^{2}(\hat{\theta}_{B\!E_k})\Delta\theta_{B\!E_k}^{4}}{8}\Big)\Big(\alpha_{uv}\sin(\hat{\theta}_{B\!E_k})\Delta\theta_{B\!E_k}\Big)\\
&\approx\frac{\alpha_{uv}\cos(\hat{\theta}_{B\!E_k})}{2}\Delta\theta_{B\!E_k}^{2}+\alpha_{uv}\sin(\hat{\theta}_{B\!E_k})\Delta\theta_{B\!E_k}.
\end{aligned}
\end{equation}
Then, combining (\ref{Upsilon_2}), (\ref{cos_2}) and (\ref{sin}), we can obtain the expression of $\Upsilon_{2k}(u,v)$ at the top of this page.

\par
Now, combining (\ref{H_BEk_pq_23}), (\ref{Gamma_1k}) and (\ref{Gamma_2k}), we can finally obtain the analytic expression of $H_{B\!E_k}(u,v)$ and the proof is completed.

\appendices
\section*{Appendix B}
The $p$-th entry of $\textbf{h}(\theta_{B\!E_k})$ can be expressed as
\renewcommand\theequation{B1}
\begin{equation}\label{h_p}
\begin{aligned}
&h_{p}(\theta_{B\!E_k})=\\
&\sqrt{\frac{g_{B\!E_k}}{N}}e^{-j2\pi\frac{(p-(N+1)/2)l}{\lambda}\cos(\hat{\theta}_{B\!E_k}+\Delta\theta_{B\!E_k})}\\
&=\sqrt{\frac{g_{B\!E_k}}{N}}e^{-j\alpha_{p}\Big(\cos(\hat{\theta}_{B\!E_k})\cos(\Delta\theta_{B\!E_k})-\sin(\hat{\theta}_{B\!E_k})\sin(\Delta\theta_{B\!E_k})\Big)},
\end{aligned}
\end{equation}
where $\alpha_{p}=\frac{2\pi(p-(N+1)/2)l}{\lambda}$.
Since $\Delta\theta_{\mathrm{max}}$ is a minimum value nears zero, we can expand the trigonometric functions of the direction angle estimation error in (\ref{h_p}) by using second-order Taylor expansion, and (\ref{h_p}) is transformed as follows
\renewcommand\theequation{B2}
\begin{equation}
\begin{aligned}
&h_{p}(\theta_{B\!E_k})=\\
&\sqrt{\frac{g_{B\!E_k}}{N}}e^{-j\alpha_{p}\Big(\cos(\hat{\theta}_{B\!E_k})(1-\frac{\Delta\theta_{B\!E_k}^{2}}{2})-\sin(\hat{\theta}_{B\!E_k})\Delta\theta_{B\!E_k}\Big)}\\
&=\!\sqrt{\frac{g_{B\!E_k}}{N}}\!e^{-j\!\alpha_{p}\!\cos(\!\hat{\theta}_{B\!E_k}\!)}\!e^{j\!\alpha_{p}\!\Big(\!\cos(\hat{\theta}_{B\!E_k})\!\frac{\Delta\theta_{B\!E_k}^{2}}{2}+\sin(\hat{\theta}_{B\!E_k})\!\Delta\theta_{B\!E_k}\!\Big)}\\
&=h_{p}(\hat{\theta}_{B\!E_k})\Big(\cos(\xi_{p}^{k})+j\sin(\xi_{p}^{k})\Big)\\
&=\Gamma_{1k}(p)+j\Gamma_{2k}(p).
\end{aligned}
\end{equation}
where
\renewcommand\theequation{B3}
\begin{equation}
h_{p}(\hat{\theta}_{B\!E_k})=\sqrt{\frac{g_{B\!E_k}}{N}}e^{-j\alpha_{p}\!\cos(\hat{\theta}_{B\!E_k})}
\end{equation}
denotes the $p$-th entry of $\textbf{h}(\hat{\theta}_{B\!E_k})$, and
\begin{equation}\label{xi_p}
\renewcommand\theequation{B4}
\xi_{p}^{k}=\alpha_{p}\Big(\cos(\hat{\theta}_{B\!E_k})\frac{\Delta\theta_{B\!E_k}^{2}}{2}+\sin(\hat{\theta}_{B\!E_k})\Delta\theta_{B\!E_k}\Big),
\end{equation}
\begin{equation}\label{gamma_1k}
\renewcommand\theequation{B5}
\Gamma_{1k}(p)=h_{p}(\hat{\theta}_{B\!E_k})\cos(\xi_{p}^{k}),
\end{equation}
\begin{equation}\label{gamma_2k}
\renewcommand\theequation{B6}
\Gamma_{2k}(p)=h_{p}(\hat{\theta}_{B\!E_k})\sin(\xi_{p}^{k}).
\end{equation}

\par
Then performing the same operation on the trigonometric functions in (B5) and (B6) as shown in APPENDIX A, we finally obtain the expression of $h_{p}(\theta_{B\!E_k})$,
\renewcommand\theequation{B7}
\begin{equation}
\begin{aligned}
&h_{p}(\theta_{B\!E_k})=\\
&h_{p}(\hat{\theta}_{B\!E_k})\Big(\cos(\xi_{p}^{k})+j\sin(\xi_{p}^{k})\Big)\\
&=h_{p}(\hat{\theta}_{B\!E_k})+j\alpha_{p}\sin(\hat{\theta}_{B\!E_k})h_{p}(\hat{\theta}_{B\!E_k})\Delta\theta_{B\!E_k}\\
&\quad+\frac{1}{2}\alpha_{p}\Big(j\cos(\hat{\theta}_{B\!E_k})-\alpha_{p}\sin^{2}(\hat{\theta}_{B\!E_k})\Big)h_{p}(\hat{\theta}_{B\!E_k})\Delta\theta_{B\!E_k}^{2}\\
&\quad-\frac{1}{4}\alpha_{p}^{2}\sin(2\hat{\theta}_{B\!E_k})h_{p}(\hat{\theta}_{B\!E_k})\Delta\theta_{B\!E_k}^{3}\\
&=h_{p}(\hat{\theta}_{B\!E_k})+\Delta h_{p}^{k},
\end{aligned}
\end{equation}
where
\renewcommand\theequation{B8}
\begin{equation}\label{Delta}
\begin{aligned}
&\Delta h_{p}^{k}=\\
&j\alpha_{p}\sin(\hat{\theta}_{B\!E_k})h_{p}(\hat{\theta}_{B\!E_k})\Delta\theta_{B\!E_k}\\
&+\frac{1}{2}\alpha_{p}\Big(j\cos(\hat{\theta}_{B\!E_k})-\alpha_{p}\sin^{2}(\hat{\theta}_{B\!E_k})\Big)h_{p}(\hat{\theta}_{B\!E_k})\Delta\theta_{B\!E_k}^{2}\\
&-\frac{1}{4}\alpha_{p}^{2}\sin(2\hat{\theta}_{B\!E_k})h_{p}(\hat{\theta}_{B\!E_k})\Delta\theta_{B\!E_k}^{3}
\end{aligned}
\end{equation}
is the $p$-th entry of the channel estimation error of the $k$-th eavesdropper.

\par
Since the direction angle estimation error satisfies $\Delta\theta_{B\!E_k}\in[-\Delta\theta_{\mathrm{max}},\Delta\theta_{\mathrm{max}}]$, and $\Delta\theta_{\mathrm{max}}$ is a minimum value nears zero, then we can omit the terms of the second-order and the third-order of $\Delta\theta_{B\!E_k}$ in (\ref{Delta}).
Therefore, $\Delta h_{p}^{k}$ can be approximately rewritten as
\renewcommand\theequation{B9}
\begin{equation}
\Delta h_{p}^{k}\approx j\alpha_{p}\sin(\hat{\theta}_{B\!E_k})h_{p}(\hat{\theta}_{B\!E_k})\Delta\theta_{B\!E_k},
\end{equation}
\par
\noindent and the modulus of $\Delta h_{p}^{k}$ is
\renewcommand\theequation{B10}
\begin{equation}
|\Delta h_{p}^{k}|= |j\alpha_{p}\sin(\hat{\theta}_{B\!E_k})h_{p}(\hat{\theta}_{B\!E_k})||\Delta\theta_{B\!E_k}|.
\end{equation}

\par
Apparently, $|\Delta h_{p}^{k}|$ is a monotonically increasing function with respect to $|\Delta\theta_{B\!E_k}|$. Therefore, $|\Delta h_{p}^{k}|$ can take the maximum value when and only when $\Delta\theta_{B\!E_k}$ equals to the maximum value $\Delta\theta_{\mathrm{max}}$ or the minimum value $-\Delta\theta_{\mathrm{max}}$, i.e.,
\renewcommand\theequation{B11}
\begin{equation}
|\Delta h_{p}^{k}|\leq |j\alpha_{p}\sin(\hat{\theta}_{B\!E_k})h_{p}(\hat{\theta}_{B\!E_k})||\Delta\theta_{\mathrm{max}}|.
\end{equation}

\par
It shows that $|\Delta h_{p}^{k}|$ is bounded, then each entry in $|\Delta \textbf{h}_{B\!E_k}|$ is also bounded, thus $|\Delta \textbf{h}_{B\!E_k}|$ is norm-bounded, which means that the channel estimation error of the eavesdroppers are norm-bounded and the proof is completed.

\par
We use $\varepsilon_k$ to represent the maximum value of the bound of the norm of the channel estimation error of the $k$-th eavesdropper, then
\renewcommand\theequation{B12}
\begin{equation}
\varepsilon_k=\sqrt{\sum_{p=1}^{N}|\alpha_{p}\sin(\hat{\theta}_{B\!E_k})h_{p}(\hat{\theta}_{B\!E_k})\Delta\theta_{\mathrm{max}}|^{2}}.
\end{equation}

\ifCLASSOPTIONcaptionsoff
  \newpage
\fi

\bibliographystyle{IEEEtran}
\bibliography{IEEEabrv,RobustDMMUMISOYang0330} %注意 ，后面不能有空格
\end{document}